\documentclass[]{spie}  

\usepackage{amsmath,amsfonts,amssymb, mathrsfs}
\usepackage{bigints}
\usepackage{subfig}
\usepackage{graphicx}
\usepackage[colorlinks=true, allcolors=blue]{hyperref}
\usepackage{stmaryrd}
\usepackage{enumerate}
\usepackage{comment}

\usepackage[braket,qm]{qcircuit}
\pdfmapfile{+sansmathaccent.map}

\usepackage{braket}
\usepackage{physics}

\usepackage{xcolor}
\usepackage{listings}
\usepackage{color}

\definecolor{codegreen}{rgb}{0,0.6,0}
\definecolor{codegray}{rgb}{0.5,0.5,0.5}
\definecolor{codepurple}{rgb}{0.58,0,0.82}
\definecolor{backcolour}{rgb}{0.95,0.95,0.92}
 
\lstdefinestyle{mystyle}{
    backgroundcolor=\color{backcolour},   
    commentstyle=\color{codegreen},
    keywordstyle=\color{magenta},
    numberstyle=\tiny\color{codegray},
    stringstyle=\color{codepurple},
    basicstyle=\footnotesize,
    breakatwhitespace=false,         
    breaklines=true,                 
    captionpos=b,                    
    keepspaces=true,                 
    numbers=left,                    
    numbersep=5pt,                  
    showspaces=false,                
    showstringspaces=false,
    showtabs=false,                  
    tabsize=2
}
\usepackage{todonotes}

\DeclareSymbolFont{largesymbolsA}{U}{txexa}{m}{n}
\DeclareMathSymbol{\varprod}{\mathop}{largesymbolsA}{16}

\title{Quantum amplitude estimation algorithms on IBM quantum devices}

\author[a]{Pooja Rao}
\author[b,*]{Kwangmin Yu}
\author[c]{Hyunkyung Lim}
\author[c]{Dasol Jin}
\author[d]{Deokkyu Choi}

\affil[a]{Department of Mathematics, Stony Brook University, Stony Brook, New York 11794, USA}
\affil[b]{Computational Science Initiative, Brookhaven National Laboratory, Upton, New York 11973, USA}
\affil[c]{Department of Applied Mathematics and Statistics, Stony Brook University, Stony Brook, New York 11794, USA}
\affil[d]{Department of Chemistry, Stony Brook University, Stony Brook, New York 11794, USA}

\authorinfo{* : Send correspondence to Kwangmin Yu,  E-mail : kyu@bnl.gov\\
}

\begin{document} 
\maketitle

\begin{abstract}
Since the publication of the Quantum Amplitude Estimation (QAE) algorithm by Brassard et al., 2002, several variations have been proposed, such as Aaronson et al., 2019, Grinko et al., 2019, and Suzuki et al., 2020. The main difference between the original and the variants is the exclusion of Quantum Phase Estimation (QPE) by the latter. This difference is notable given that QPE is the key component of original QAE, but is composed of many operations considered expensive for the current NISQ era devices. We compare two recently proposed variants (Grinko et al., 2019 and Suzuki et al., 2020) by implementing them on the IBM Quantum device using Qiskit, an open source framework for quantum computing. We analyze and discuss advantages of each algorithm from the point of view of their implementation and performance on a quantum computer.
\end{abstract}


\section{Introduction}
\label{sec:introduction}

The field of quantum computing, although still in its nascent phase, has seen some significant development since its inception in the early 1980s. Companies such as IBM have ushered in a new era of quantum computing by allowing public access to their quantum computers. However, quantum noise is still a major obstacle for today's quantum machines, often referred to as the Noisy Intermediate-Scale Quantum (NISQ) devices~\cite{Preskill2018quantumcomputingin}. Thus, it is important to design quantum computing algorithms that work around the limitations of the current NISQ devices.
Among the quantum computing algorithms that exhibit quantum speedup over their classical counterparts, Grover's search~\cite{grover1996fast} and its generalization, quantum amplitude estimation~\cite{brassard2002quantum}(QAE), have special importance because of the range of applications they cover. However, the foundation of original QAE is Quantum Phase Estimation (QPE), which requires a large number of controlled operations, making it largely infeasible on NISQ devices. To mitigate this issue, a number of improvements have been proposed in recent years that require appreciably less number of qubits and quantum circuit depth~\cite{grinko2019iterative, suzuki2020amplitude, aaronson2020quantum, wie2019simpler, nakaji2020faster}, thus making them more appropriate from a practical standpoint~\cite{Yu2020Practical,Yu2020Comparison}. The common thread that runs amongst most of these recent QAE algorithms is their exclusion of QPE, which makes them better suited for the near term quantum devices.

One important application of quantum amplitude estimation is computing integrals using Monte Carlo integration (MCI). In this study, we compute the integral of the function $\sin^2 x$ on the interval $[0, \pi/4]$ using two recent QAE variants. We use $2$ and $3$-qubit domains on IBM Q Vigo and $2$, $3$ and $10$-qubit domains on IBM Q Simulator, using Qiskit~\cite{qiskit}. The first algorithm by Suzuki et al.~\cite{suzuki2020amplitude}, known as the maximum likelihood quantum amplitude estimation (MLQAE), uses the maximum likelihood estimation on the measurements made on a sequence of quantum circuits, while the second algorithm by Grinko et al.~\cite{grinko2019iterative}, known as the iterative quantum amplitude estimation (IQAE), uses an iterative optimization of QAE. We compare these two QAE algorithms from the perspective of accuracy and efficiency for quantum Monte Carlo integration. Additionally, by comparing the quantum device runs with the simulator runs for the $2$ and $3$-qubit domains, we can isolate the effect of quantum device error on the performance of the algorithms.

\section{Notations}
In the following sections, Dirac's bra-ket notation will be used for qubit representation and arithmetic, such as the tensor product.
For a multi-qubit system, consecutive binary number strings have the most significant qubit located on the left and the least significant qubit on the right.
For example, we have $\ket{0} \otimes \ket{1} \otimes \ket{1} = \ket{0} \ket{1} \ket{1} = \ket{011}$ in the binary representation, and $\ket{011} = \ket{3}$ and $\ket{110} = \ket{6}$ in the decimal representation of the computational basis.
In the decimal representation, the number of qubits, $n$, is denoted by a subscript as in $\ket{0}_n$.
The dimension of a square matrix is also denoted by a subscript.
For example, $\mathbb{I}_{n}$ denotes the $n\times n$ identity matrix.

In quantum circuit diagrams, the top and the bottom qubits represent the least and the most significant qubits, respectively. 
The Hadamard matrix, H = $\frac{1}{\sqrt{2}} \begin{pmatrix} 1 & 1\\ 1 & -1\end{pmatrix}$, is a unitary matrix that is also Hermitian, so it is its own inverse.
The three Pauli matrices (Pauli gates when they are used in quantum circuits) $X$, $Y$ and $Z$ are,

\begin{equation*}
\label{eq:pauli}
    X = \begin{pmatrix} 0 & 1\\ 1 & 0\end{pmatrix}, ~~
    Y = \begin{pmatrix} 0 & -i\\ i & 0\end{pmatrix}, ~~
  and~~  Z = \begin{pmatrix} 1 & 0\\ 0 & -1\end{pmatrix}.
\end{equation*}

\section{Quantum Amplitude Amplification and Estimation}
\label{sec:qae}


In this section, we briefly review quantum amplitude amplification (QAA) and quantum amplitude estimation algorithms  \cite{brassard2002quantum}. Quantum amplitude amplification is a generalization of Grover's search algorithm \cite{grover1996fast}, which maintains the quadratic quantum speedup offered by Grover's search.
While the original Grover's algorithm searches for one solution in the given domain, QAA searches for multiple solutions.
Thus, when counting the number of solutions is of more interest than an individual solution, QAE can be applied to estimate the number of solutions.
QAA and QAE are the fundamental building blocks for the quantum implementation of Monte Carlo integration \cite{Yu2020Practical}.
Montanaro~\cite{montanaro2015montecarlo} has shown how QAA leads to a quadratic speed up for general Monte Carlo integration. 

Suppose we have an $n$-qubit problem domain, $\mathscr{D}$, such that $| \mathscr{D} | = N = 2^n$, and a unitary operator $\mathcal{A}$ acting on $n+1$ qubits such that
\begin{equation}
\label{eq:qae_a}
    \ket{\Psi} = \mathcal{A} \ket{0}_n \ket{0} = \sqrt{1-a} \ket{\psi_0}_n \ket{0} + \sqrt{a} \ket{\psi_1}_n \ket{1},
\end{equation}
\noindent where the good state is $\ket{\psi_1}_n$ and the bad state is $\ket{\psi_0}_n$.
The main goal of QAA and QAE is to estimate $a$, the probability of measuring the good state.
To achieve the quantum speedup, instead of measuring the last qubit of $\ket{\Psi} = \mathcal{A} \ket{0}_n \ket{0}$ directly, $\ket{\Psi}$ is first amplified by the unitary operator,

\begin{equation}
\label{eq:qae_def_q}
    \textbf{Q} = \mathcal{A} \textbf{S}_0 \mathcal{A}^{-1} \textbf{S}_{\chi},
\end{equation}

\noindent where $\textbf{S}_0 = \mathbb{I}_{n+1} - 2 \ket{0}_{n+1} \bra{0}_{n+1}$ and
$\textbf{S}_{\chi} = ( \bigotimes\limits^{n} \mathbb{I}_2 ) \otimes Z$.
$\textbf{S}_{\chi}$ puts a negative sign to the good state, $\ket{\psi_1}_n \ket{1}$, and does nothing to the bad state, $\ket{\psi_0}_n \ket{0}$.
Let us define a parameter $\theta \in [0, \pi/2]$ so that $\sin^2 \theta = a$. With this, we can rewrite Eq.~(\ref{eq:qae_a}) as
\begin{equation}
\label{eq:qae_theta}
    \ket{\Psi} = \mathcal{A} \ket{0}_n \ket{0} = \cos \theta \ket{\psi_0}_n \ket{0} + \sin \theta \ket{\psi_1}_n \ket{1}.
\end{equation}

\noindent By applying $\textbf{Q}$ (amplitude amplification operator) $m$ times repeatedly on $\ket{\Psi}$, we get 

\begin{equation}
\label{eq:qae_qm}
    \textbf{Q}^m \ket{\Psi} = \cos ((2m+1) \theta) \ket{\psi_0} \ket{0} + \sin ((2m+1) \theta) \ket{\psi_1} \ket{1}.
\end{equation}

\noindent From Eqs.~(\ref{eq:qae_a}) and (\ref{eq:qae_theta}), it can be observed that the measurement, after applying $\textbf{Q}^m$ on $\mathcal{A} \ket{0}_n \ket{0}$, shows a quadratically larger probability of obtaining the good state than measuring $\mathcal{A} \ket{0}_n \ket{0}$ directly \cite{brassard2002quantum}, provided $\theta$ is sufficiently small so that $ (2m+1) \theta < \frac{\pi}{2}$.


The canonical QAE~\cite{brassard2002quantum} algorithm estimates $\theta$ in Eq.~(\ref{eq:qae_theta}) by QPE, which includes the inverse quantum Fourier transform. QPE is implemented by the controlled operation on the $\textbf{Q}$ operator. 
Hence, it needs a number of multi-controlled operations, which are further decomposed into many basis gates, increasing the circuit depth.
When this algorithm is implemented on NISQ devices, the accuracy is strongly limited by the lack of direct connectivity between qubits because all the ancilla registers need connectivity with the target register and a sufficiently large number of ancillae are needed to ensure desired accuracy of the estimation.   

On the other hand, MLQAE~\cite{suzuki2020amplitude} is implemented by post-processing the results of the quantum computation (and measuring the process without QPE) using maximum likelihood estimation. Since different levels of amplification (depending on power of $\textbf{Q}$) can be executed independently, the algorithm is parallelizable. 
IQAE~\cite{grinko2019iterative} also post-processes results from quantum circuit runs. In the post-processing, it estimates the optimal power $k$ of $\textbf{Q}$ and applies $\textbf{Q}^k$ to $\ket{\Psi}$ (see Eq.~(\ref{eq:qae_a})) in the next iteration. The iteration continues until the specified error bound is met.
The main benefit of MLQAE and IQAE over the canonical QAE is that they do not need extra ancilla qubits to read out the amplitude, $\sqrt{a}$, of Eq.~(\ref{eq:qae_a}), except to read out the least significant qubit.
The analysis of error bounds and the number of queries (application of $\mathcal{A}$ and $\mathcal{A}^{-1}$) of MLQAE and IQAE are discussed in the literature \cite{suzuki2020amplitude, grinko2019iterative, aaronson2020quantum}.
Both MLQAE and IQAE need $O \bigl( \frac{1}{\epsilon} \bigl( \sqrt{\frac{K}{N}} \bigr)  \bigr)$ oracle queries to estimate $K$ good states in $N$ samples (domain) with error $\epsilon$ \cite{aaronson2020quantum}.
The implementation aspects and results obtained from executing MLQAE and IQAE on IBM quantum simulator and IBM quantum devices will be discussed in the following sections.



\section{Numerical Integration}
\label{sec:integration}

\begin{figure*}[t]
\centering
  \subfloat[Four subintervals (discretization)]{
    \includegraphics[scale=0.45]{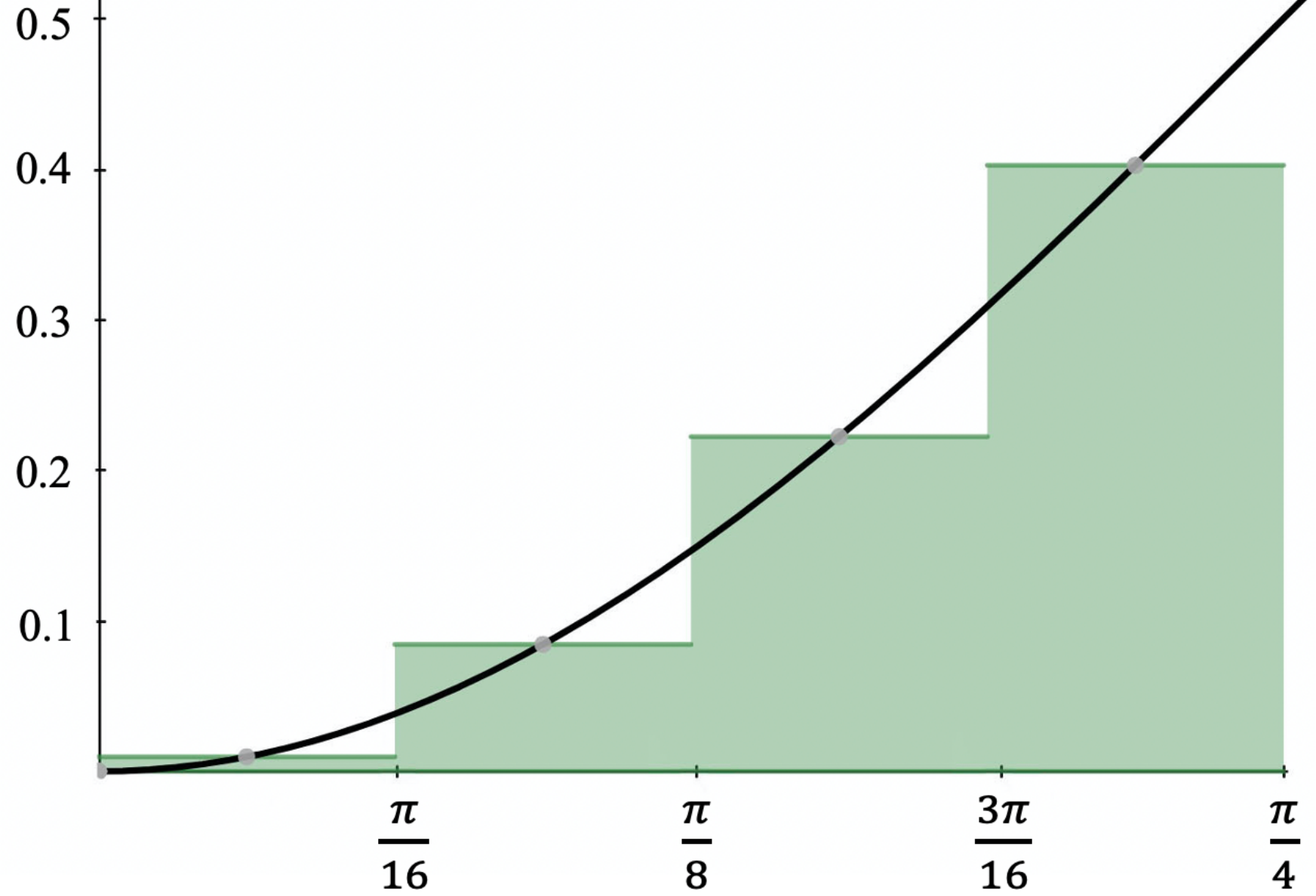}}
  \qquad
  \subfloat[Eight subintervals (discretization)]{
    \includegraphics[scale=0.45]{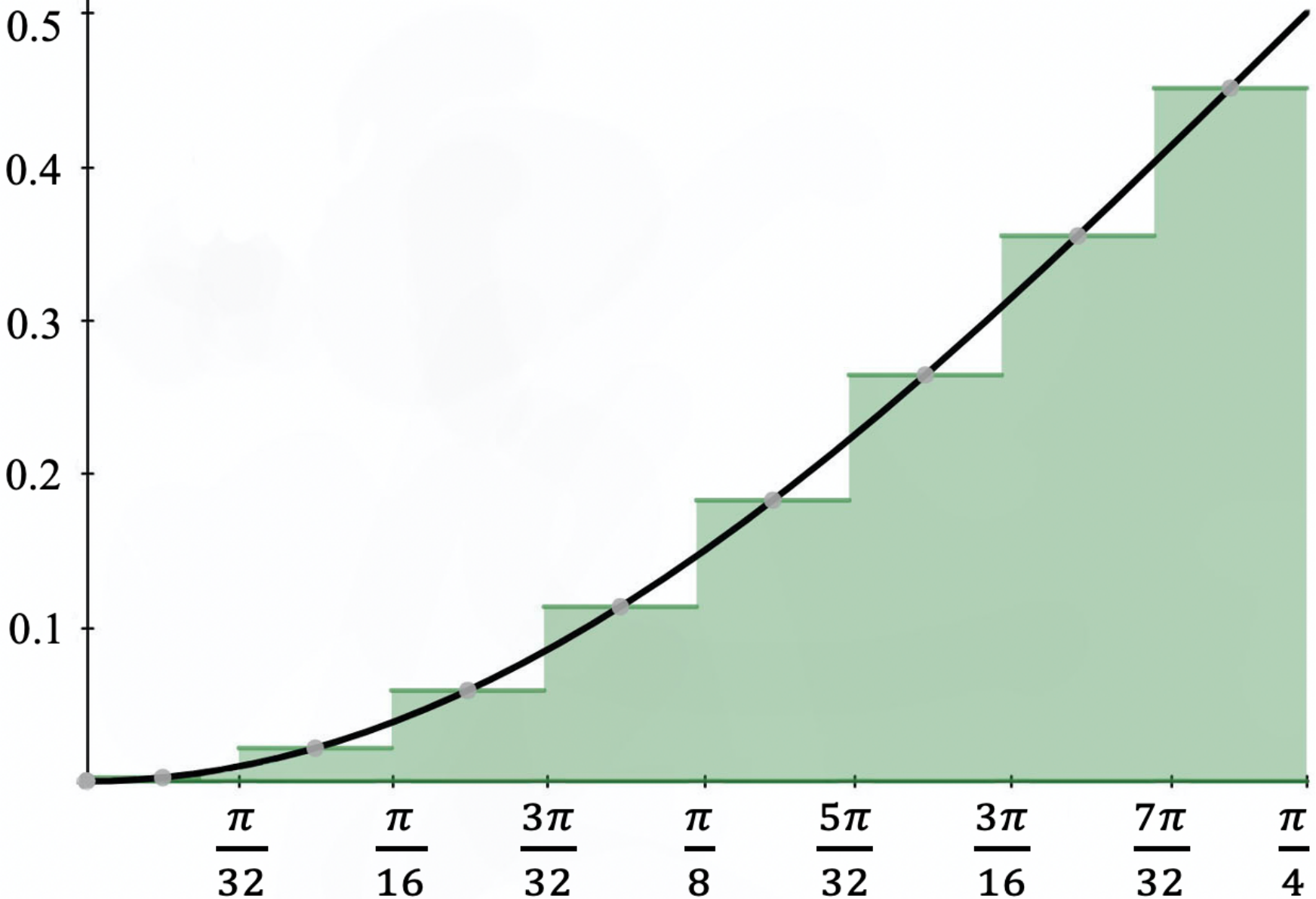}
  }\\
  \vspace*{5mm}
  \caption{The Riemann sums using the midpoint in each subinterval for $\sin^2 x$ on $[0, \pi/4]$}
  \label{fig:Riemann_sum}
\end{figure*}

In this section, we briefly discuss the numerical integration we use to assess the performance of the two QAE algorithms.
If $f$ is a function of $x$ defined on $a \leq x \leq b$, we divide the interval $\left[a, b \right]$ into $N$ subintervals of equal width $\Delta x = (b - a)/N$.
We let $x_0 (= a), x_1, x_2, \cdots, x_N (=b)$ be the end points of the subintervals and let $x_0^* , x_1^* , x_{N-1}^*$ be any sample points in the subintervals.
That is, we have $x_i^* \in \left[x_i, x_{i+1} \right]$.
Then the definite integral of $f$ from $a$ to $b$ is,
\begin{equation}
\label{eq:def_integral}
\int_{a}^{b}  f(x) ~ dx =  \lim_{N \rightarrow \infty}  \sum_{i=0}^{N-1} f(x_i^*) \Delta x,
\end{equation}
if the limit exists. 
When the limit does exist, we say that $f$ is integrable on $\left[a, b \right]$ \cite{stewart2006calculus}.
It is well-known that $f$ is integrable on $\left[a, b \right]$ when $f$ is continuous on $\left[a, b \right]$.
If the limit does exist, it does not depend on where in the subinterval the sample points, $x_i^*$, have been chosen from. Therefore, our choice of midpoints of the subinterval as the sample points does not influence the convergence.
The approximate value of a definite integral is computed using only a finite number of subintervals,
\begin{equation}
\label{eq:def_integral_approx}
\lim_{N \rightarrow \infty}  \sum_{i=0}^{N-1} f(x_i^*) \Delta x \approx \sum_{i=0}^{N-1}  f(x_i^*) \frac{(b-a)}{N}, 
\end{equation}
where $N$ is the number of subintervals or samples.
For an $n$-qubit computation domain, there are $N = 2^n$ subintervals.


Figures~\ref{fig:Riemann_sum}(a) and (b) show the midpoint Riemann sums of four and eight subintervals, respectively.
The area in Fig.~\ref{fig:Riemann_sum}(a) is
\begin{align}
\label{eq:int_pi_over_4_2q_expand}
\notag \int_{0}^{\pi/4}  \sin^2 x ~ dx \approx &\sum_{i=0}^{3}  \sin^2 (x_i^*) \frac{(\pi/4)}{4} = \frac{\pi}{4} \sum_{i=0}^{3}  \sin^2 (x_i^*) \frac{1}{4} = 
\frac{\pi}{4} \sum_{i=0}^{3} \sin^2 \left( \frac{(i + 1/2)}{4} \frac{\pi}{4}  \right) \cdot \frac{1}{4}=\\
& \frac{\pi}{4} \cdot \left( \sin^2 \left( \frac{1}{8} \cdot \frac{\pi}{4} \right) + \sin^2 \left( \frac{3}{8} \cdot \frac{\pi}{4} \right) + \sin^2 \left( \frac{5}{8} \cdot \frac{\pi}{4} \right) + \sin^2 \left( \frac{7}{8} \cdot \frac{\pi}{4} \right) \right) \cdot \frac{1}{4} \approx 0.141085, 
\end{align}
and the area in Fig.~\ref{fig:Riemann_sum}(b) is
\begin{align}
\label{eq:int_pi_over_4_3q_expand}
\notag \int_{0}^{\pi/4}  \sin^2 x ~ dx \approx &\sum_{i=0}^{7}  \sin^2 (x_i^*) \frac{(\pi/4)}{8} = \frac{\pi}{8} \sum_{i=0}^{7}  \sin^2 (x_i^*) \cdot \frac{1}{8} = 
\frac{\pi}{4} \sum_{i=0}^{7} \sin^2 \left( \frac{(i + 1/2)}{8} \frac{\pi}{4}  \right) \cdot \frac{1}{8} = \\
& \frac{\pi}{4} \cdot \left( \sin^2 \left( \frac{1}{16} \cdot \frac{\pi}{4} \right) + \sin^2 \left( \frac{3}{16} \cdot \frac{\pi}{4} \right) + \sin^2 \left( \frac{5}{16} \cdot \frac{\pi}{4} \right) + \cdots + \sin^2 \left( \frac{15}{16} \cdot \frac{\pi}{4} \right) \right) \cdot \frac{1}{8} \approx 0.142297, 
\end{align}
where $x_i^*$ is the midpoint of each subinterval.
On the other hand, the exact value of the definite integral is
\begin{equation}
\label{eq:int_pi_over_4}
\int_{0}^{\pi/4}  \sin^2 x ~ dx = \frac{\pi}{8} - \frac{1}{4} \approx 0.142699.
\end{equation}
Equations~(\ref{eq:int_pi_over_4_2q_expand}) and (\ref{eq:int_pi_over_4_3q_expand}) demonstrate how the Riemann sums approach the exact value as more subintervals are considered.

\section{Implementation}
\label{sec:implementation}

In this section, we formulate the operator $\mathcal{A}$ used in Eqs.~(\ref{eq:qae_a}), (\ref{eq:qae_def_q}), and (\ref{eq:qae_theta}) for the numerical integration discussed in Sec.~\ref{sec:integration}, $\int_{0}^{\pi/4}  \sin^2 x ~ dx$.

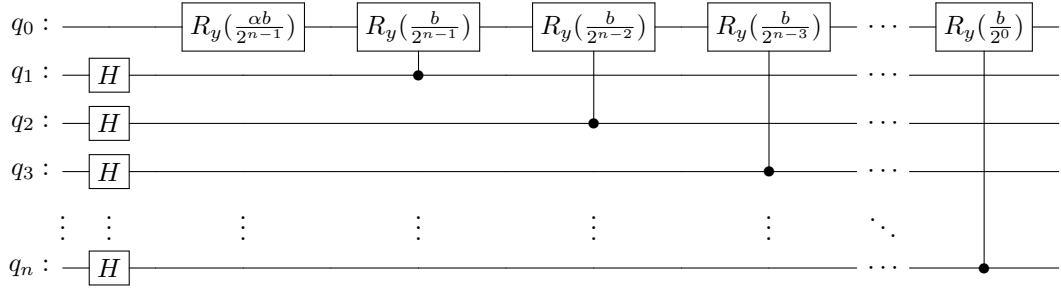
\begin{figure}[t!]
\[ 
\Qcircuit @C=1.0em @R=0.0em @!R {
    \lstick{ {q}_{0} :  } & \qw & \qw & \gate{R_y(\frac{\alpha b}{2^{n-1}})} & \qw & \gate{R_y(\frac{b}{2^{n-1}})} & \qw & \gate{R_y(\frac{b}{2^{n-2}})} & \qw & \gate{R_y(\frac{b}{2^{n-3}})} & \qw & \cdots &  & \gate{R_y(\frac{b}{2^{0}})} & \qw  \\
    \lstick{ {q}_{1} :  } & \gate{H} & \qw & \qw & \qw & \ctrl{-1} & \qw & \qw & \qw & \qw & \qw & \cdots &  & \qw & \qw \\
    \lstick{ {q}_{2} :  } & \gate{H} & \qw & \qw & \qw & \qw & \qw & \ctrl{-2} & \qw & \qw & \qw & \cdots &  & \qw & \qw \\
    \lstick{ {q}_{3} :  } & \gate{H} & \qw & \qw & \qw & \qw & \qw & \qw & \qw & \ctrl{-3} & \qw & \cdots &  & \qw & \qw \\
    \vdots & \vdots & & \vdots & & \vdots & & \vdots & & \vdots &  & \ddots & &  & &  \\
    \lstick{ {q}_{n} :  } & \gate{H} & \qw & \qw & \qw & \qw & \qw & \qw & \qw & \qw & \qw  & \cdots &  &  \ctrl{-5} & \qw \\
}
\]
\vspace*{5mm}
\caption{The quantum circuit implementation of operator $\mathcal{A}$ in Eqs.~(\ref{eq:qae_a}), (\ref{eq:qae_theta}), and (\ref{eq:qae_func}) for $\int_{0}^{b}  f(x) ~ dx$.}
\label{fig:circuit_nq}
\end{figure} 
From Eq.~(\ref{eq:qae_a}), we derive,
\begin{equation}
\label{eq:qae_func}
    \ket{\Psi} = \mathcal{A} \ket{0}_n \ket{0} = \sum_{i=0}^{2^n - 1} \frac{1}{\sqrt{2^n}} \ket{i}_n \left( \sqrt{1-f(x_i)} \ket{0} + \sqrt{f(x_i)} \ket{1} \right),
\end{equation}
to express the function of an integral.
When we measure the least significant qubit, the measurement possibility of  $\ket{1}$ is,

\begin{equation}
\label{eq:qae_func_measure}
    \sum_{i=0}^{2^n - 1} \frac{1}{2^n} f(x_i) = \sum_{i=0}^{N - 1} f(x_i^{*}) \frac{1}{N},
\end{equation} 
where $x_i^{*}$ is some point in the $i$-th subinterval and $N = 2^n$.
Then Eq.~(\ref{eq:qae_func_measure}) multiplied by the integral interval is a Riemann sum as in Eq.~(\ref{eq:def_integral_approx}).

\begin{figure}[t!]
\[ 
\Qcircuit @C=1.0em @R=0.0em @!R {
    \lstick{ {q}_{0} :  } & \qw & \qw & \gate{R_y(\frac{\pi}{16})} & \qw & \gate{R_y(\frac{\pi}{8})} & \qw & \gate{R_y(\frac{\pi}{4})} & \qw & \qw \\
    \lstick{ {q}_{1} :  } & \gate{H} & \qw & \qw & \qw & \ctrl{-1} & \qw & \qw & \qw & \qw \\
    \lstick{ {q}_{2} :  } & \gate{H} & \qw & \qw & \qw & \qw & \qw & \ctrl{-2} & \qw & \qw \\
}
\]
\vspace{3mm}
\caption{The quantum circuit for the midpoint Riemann sums with four subintervals for $\int_{0}^{\pi/4}  \sin^2 x ~ dx$.}
\label{fig:circuit_2q}
\end{figure}
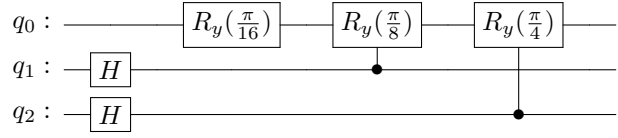

Figure~\ref{fig:circuit_nq} shows the quantum circuit implementation of operator $\mathcal{A}$ from Eqs.~(\ref{eq:qae_a}), (\ref{eq:qae_theta}), and (\ref{eq:qae_func}) for $\int_{0}^{b}  f(x) ~ dx$.
The scalar $\alpha$ ($\in [0, 1]$) in the first $R_y$ gate on the $q_0$ qubit represents the position in each subinterval as described in Eqs.~(\ref{eq:def_integral}), (\ref{eq:def_integral_approx}), and  (\ref{eq:qae_func_measure}).
For example, the Riemann sum in Fig.~\ref{fig:circuit_nq} is left, midpoint, and right Riemann sums when we have $\alpha = 0$, $0.5$, and $1$, respectively.
Figure \ref{fig:circuit_2q} shows the quantum circuit for the midpoint Riemann sums with four subintervals for $\int_{0}^{\pi/4}  \sin^2 x ~ dx$. 
The Riemann sum is visualized in Fig.~\ref{fig:Riemann_sum}(a).
The quantum states after implementing the circuit is described as,
\begin{align}
\label{eq:Riemann_sum_2q}
\notag \ket{00}\ket{0} \Rightarrow  ~
&\frac{1}{2} \ket{00} \otimes \Big( \cos (\frac{1}{8} \cdot \frac{\pi}{4}) \ket{0} + \sin (\frac{1}{8} \cdot \frac{\pi}{4}) \ket{1}  \Big) +
\frac{1}{2} \ket{01} \otimes \Big( \cos (\frac{3}{8} \cdot \frac{\pi}{4}) \ket{0} + \sin (\frac{3}{8} \cdot \frac{\pi}{4}) \ket{1}  \Big) \\
+ 
&\frac{1}{2} \ket{10} \otimes \Big( \cos (\frac{5}{8} \cdot \frac{\pi}{4}) \ket{0} + \sin (\frac{5}{8} \cdot \frac{\pi}{4}) \ket{1}  \Big) +
\frac{1}{2} \ket{11} \otimes \Big( \cos (\frac{7}{8} \cdot \frac{\pi}{4}) \ket{0} + \sin (\frac{7}{8} \cdot \frac{\pi}{4}) \ket{1}  \Big).
\end{align}
When the quantum state is measured on the least significant qubit, the measurement probability of $\ket{1}$ is
\begin{align}
\label{eq:int_pi_over_4_2q_compute}
\notag &\sum_{i=0}^{3} \sin^2 \left( \frac{(x_i + 1/2)}{4} \frac{\pi}{4}  \right) \cdot \frac{1}{4} =\\
&\left( \sin^2 \left( \frac{1}{8} \cdot \frac{\pi}{4} \right) + \sin^2 \left( \frac{3}{8} \cdot \frac{\pi}{4} \right) + \sin^2 \left( \frac{5}{8} \cdot \frac{\pi}{4} \right) + \sin^2 \left( \frac{7}{8} \cdot \frac{\pi}{4} \right) \right) \cdot \frac{1}{4} = 0.179636, 
\end{align}
which is the value divided by $\pi / 4$ (integral interval) from Eq.~(\ref{eq:int_pi_over_4_2q_expand}).
In the same manner, when we extend the circuit in Fig. \ref{fig:circuit_2q} to three qubits, the measurement probability of the least significant qubit is
\begin{align}
\label{eq:int_pi_over_4_3q_compute}
\notag &\sum_{i=0}^{7} \sin^2 \left( \frac{(i + 1/2)}{8} \frac{\pi}{4}  \right) \cdot \frac{1}{8} =\\
&\left( \sin^2 \left( \frac{1}{16} \cdot \frac{\pi}{4} \right) + \sin^2 \left( \frac{3}{16} \cdot \frac{\pi}{4} \right) + \sin^2 \left( \frac{5}{16} \cdot \frac{\pi}{4} \right) + \cdots + \sin^2 \left( \frac{15}{16} \cdot \frac{\pi}{4} \right) \right) \cdot \frac{1}{8} = 0.181178, 
\end{align}
which is the value divided by $\pi / 4$ (integral interval) from Eq.~(\ref{eq:int_pi_over_4_3q_expand}).

\vspace{3mm}
\begin{figure}[t]
\centering
  \subfloat[$\mathcal{A}$]{%
    \includegraphics[width=.43\textwidth]{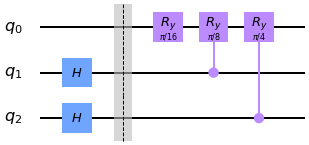}}
  \qquad
  \subfloat[$\mathcal{A}^{-1}$]{%
    \includegraphics[width=.43\textwidth]{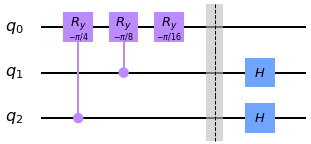}
  }\\
  \vspace{3mm}
  \caption{Quantum circuit implementation of $\mathcal{A}$ and $\mathcal{A}^{-1}$ from Eq.~(\ref{eq:qae_func}) in Qiskit.}\label{fig:A}
\end{figure}

\begin{figure*} [ht]
\begin{center}
\includegraphics[width=0.99\linewidth]{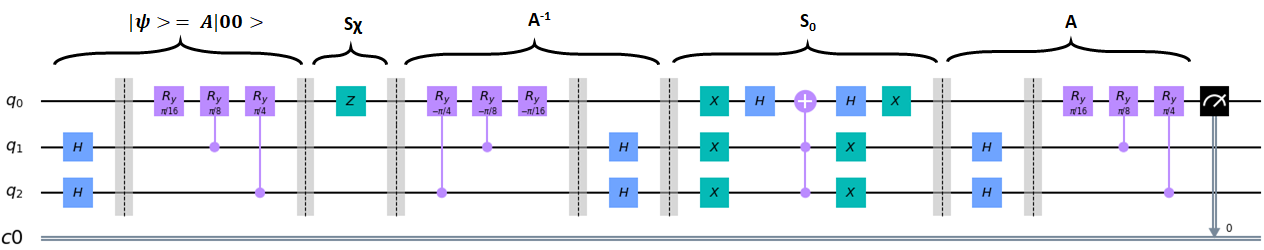}
\end{center}
\caption{Quantum circuit implementation of $\mathcal{A}$ and $\textbf{Q}$ of $2$-qubit domain from Eqs.~(\ref{eq:qae_a}) and~(\ref{eq:qae_def_q}), in Qiskit.}
\label{fig:Q} 
\end{figure*}

Therefore, the quantum state after the quantum circuit in Fig.~\ref{fig:circuit_nq} is $\ket{\Psi}$ in Eq.~(\ref{eq:qae_func}) (and Eq.~(\ref{eq:qae_a})).
Figure~\ref{fig:A} shows the implementation of $\mathcal{A}$ and $\mathcal{A}^{-1}$ for $\int_{0}^{\pi/4}  \sin^2 x ~ dx$ in two qubits.
The operator $\textbf{Q}$ in Eq.~(\ref{eq:qae_def_q}) is applied on $\mathcal{A}$ (in Fig.~\ref{fig:A}) as depicted in Fig.~\ref{fig:Q}. 
Both MLQAE and IQAE run $\textbf{Q}^k \mathcal{A}$ circuits repeatedly.
Figure \ref{fig:Q} shows the case of $k=1$ and $\textbf{Q}^k \mathcal{A}$ is implemented by applying $\textbf{Q}$ $k$-times on $\mathcal{A}$.
The main difference lies in how each method chooses the power, $k$, of $\textbf{Q}$ and what specific post-processing technique it employs.

\subsection{Maximum Likelihood Quantum Amplitude Estimation}

The key idea of MLQAE is post-processing the measurements of the quantum circuits, $\textbf{Q}^k \mathcal{A}$, with different values of $k$.
Suzuki et al.~\cite{suzuki2020amplitude} discuss two options for the circuit sequence in MLQAE, linearly incremental sequence (LIS) and exponential incremental sequence (EIS), and suggest EIS as the asymptotically optimal choice.
Thus, we use EIS, which has exponential powers of $\textbf{Q}$ from the second circuit onwards. In EIS, the $n$-th circuit has $\textbf{Q}^{2^{n-2}}$ operator applied to it following the application of $\mathcal{A}$ for $n > 1$.
For instance, when four circuits are considered for MLQAE, they are as follows: $\mathcal{A}$, $\textbf{Q} \mathcal{A}$, $\textbf{Q}^2 \mathcal{A}$, and $\textbf{Q}^4 \mathcal{A}$.
In the absence of quantum noise, more circuits would increase the accuracy of the estimation of the amplitude.


The post-processing is conducted by a maximum likelihood estimation from the measurements.
The likelihood functions and the resultant maximum likelihood function are defined on the domain $[0, \pi/2]$ for $\theta$ as
\begin{equation}
\label{eq:lfunc}
    L_k (h_k ; \theta) = \lbrace \sin^2 ((2m_k+1) \theta) \rbrace^{h_k} \lbrace \cos^2 ((2m_k+1) \theta), \rbrace^{N - h_k},
\end{equation}
and
\begin{equation}
\label{eq:mlfunc}
    L (\vec{\textbf{h}} ; \theta) = \prod\limits_{k=0}^{m}  L_k (h_k ; \theta),
\end{equation}

\noindent where $m_k$, $h_k$, and $N$ are the power of $\textbf{Q}$, hit count of $1$, and number of shots for the $k$-th circuit, respectively, and $\vec{\textbf{h}} = (h_0, h_1, \cdots, h_m)$.
The maximum likelihood estimation approximates $\hat{\theta}$, which maximizes $L (\vec{\textbf{h}} ; \hat{\theta})$ in the given domain.
But instead of $L (\vec{\textbf{h}} ; \theta)$, $\log L (\vec{\textbf{h}} ; \theta)$ is used to estimate $\hat{\theta}$ since it is a monotonically increasing, so it has the maxima in exactly the same places where the likelihood has the maxima. As a result both methods give the same answer, but the log function provides a much simpler form.

\subsection{Iterative Quantum Amplitude Estimation}

IQAE \cite{grinko2019iterative} also utilizes post-processing to estimate the optimal power $k$ of $\textbf{Q}$ at each iteration.
It starts from $k=0$ ($\ket{\Psi} = \mathcal{A} \ket{0}_n \ket{0}$) with confidence interval $[\theta_l, \theta_u] \subseteq [0, \pi/2]$, where $\theta_u$ and $\theta_l$ are the upper and lower bounds of $\theta$, respectively, in Eqs.~(\ref{eq:qae_theta}) and~(\ref{eq:qae_qm}).
The post-processing searches the maximum $k$ which puts $[(4k + 2) \theta_l, (4k + 2) \theta_u]$ in $[0, \pi]$ (upper half-plane) or $[\pi, 2\pi]$ (lower half-plane).
This method derives from the fact that the measurement probability of the good state in Eq.~(\ref{eq:qae_qm}) is $\sin^2 ((2k+1) \theta) = \{1 - \cos ((4k+2) \theta) \} / 2$ when the power of $\textbf{Q}$ is $k$ and that a cosine function is monotonic in the upper half-plane and the lower half-plane.

\section{Results}

\begin{figure*}[t]
\centering
  \subfloat[MLQAE]{
    \includegraphics[scale=0.5]{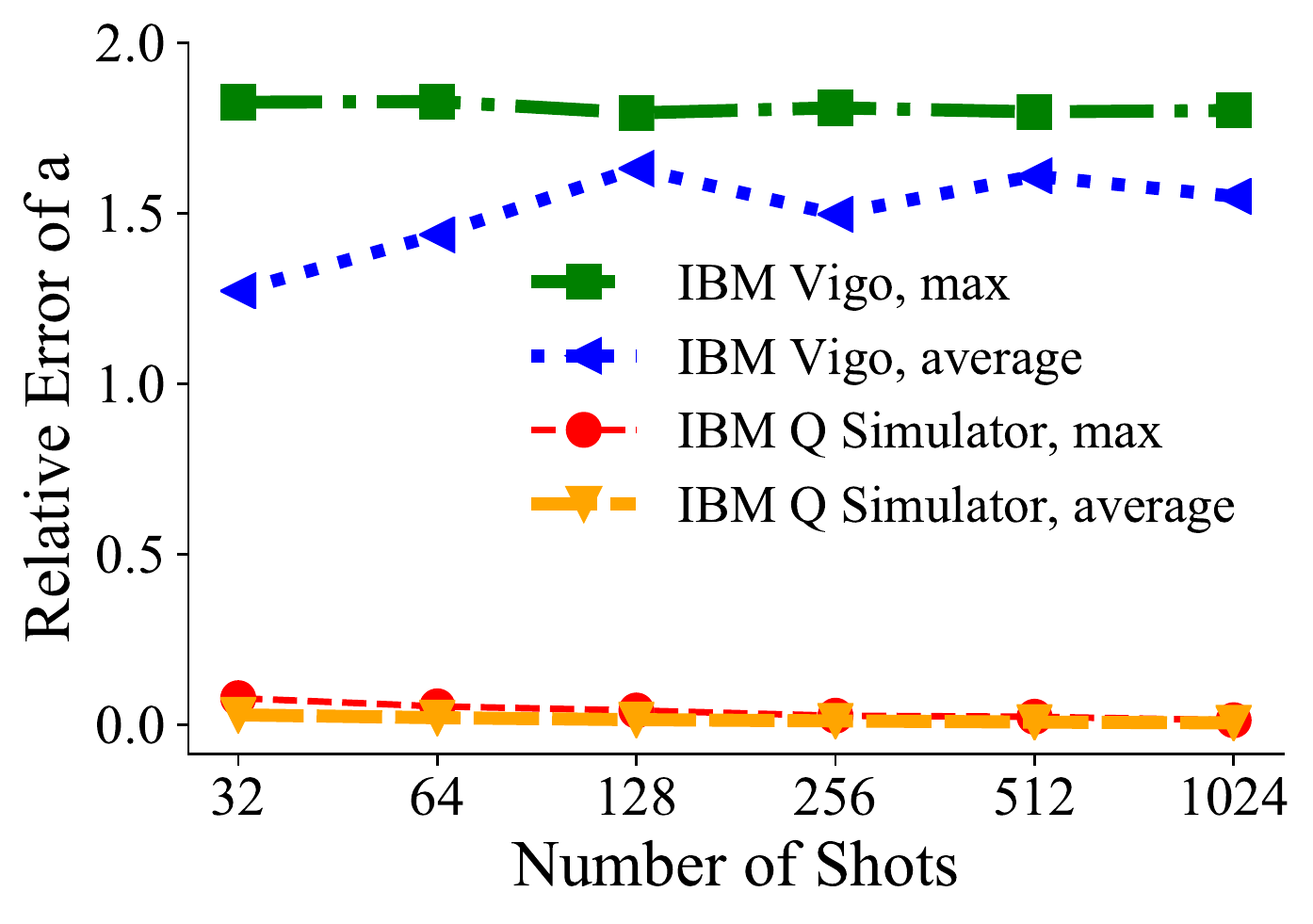}}
  \qquad
  \subfloat[IQAE]{
    \includegraphics[scale=0.5]{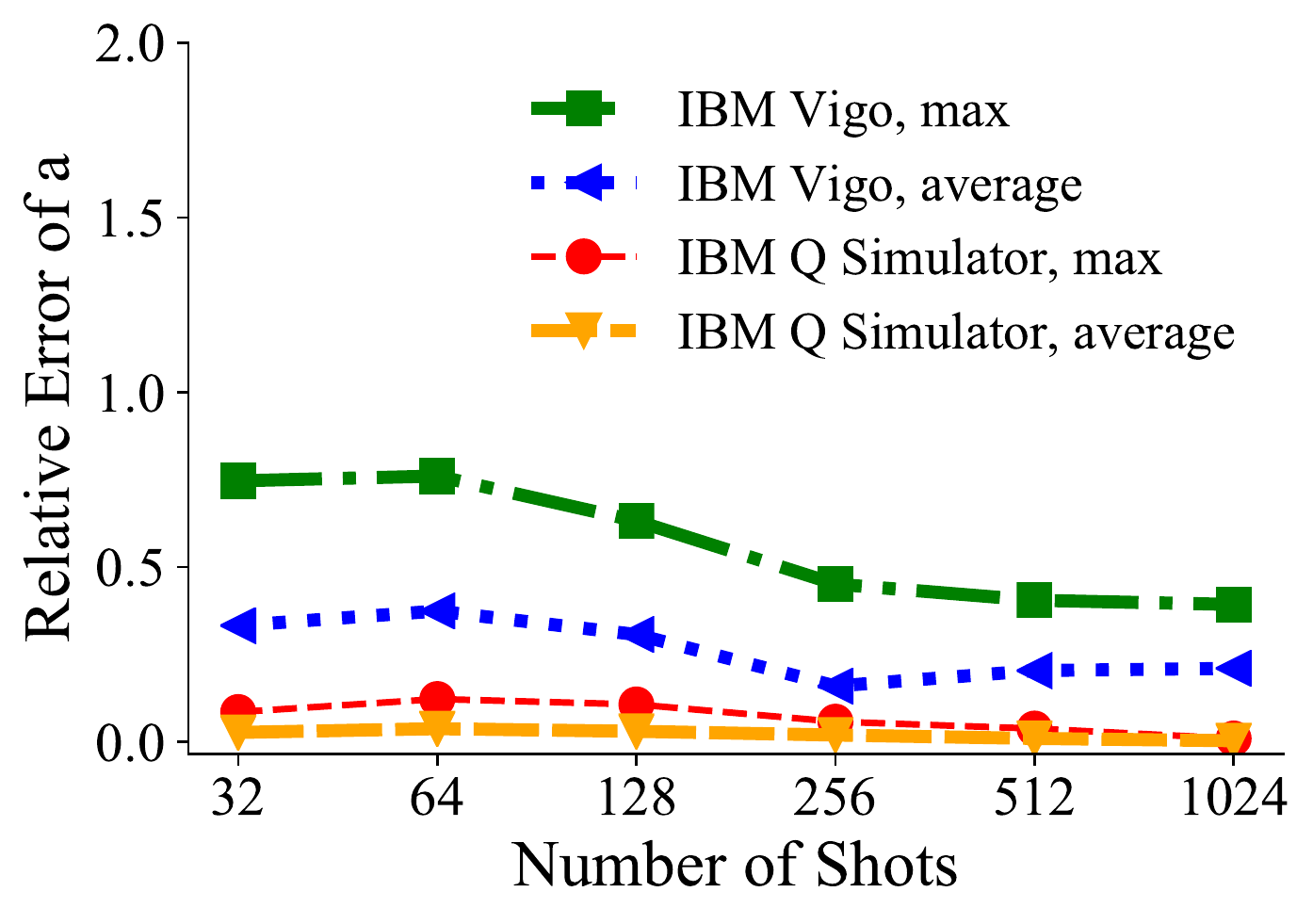}
  }\\
  \vspace{3mm}
  \caption{Relative error for $n=2$-qubit domain for MLQAE and IQAE on IBM Q Vigo and the simulator.}
  \label{fig:MLAE_IAE_rel_a_2q}
\end{figure*}

\begin{figure*}[t]
\centering
  \subfloat[MLQAE]{
    \includegraphics[scale=0.5]{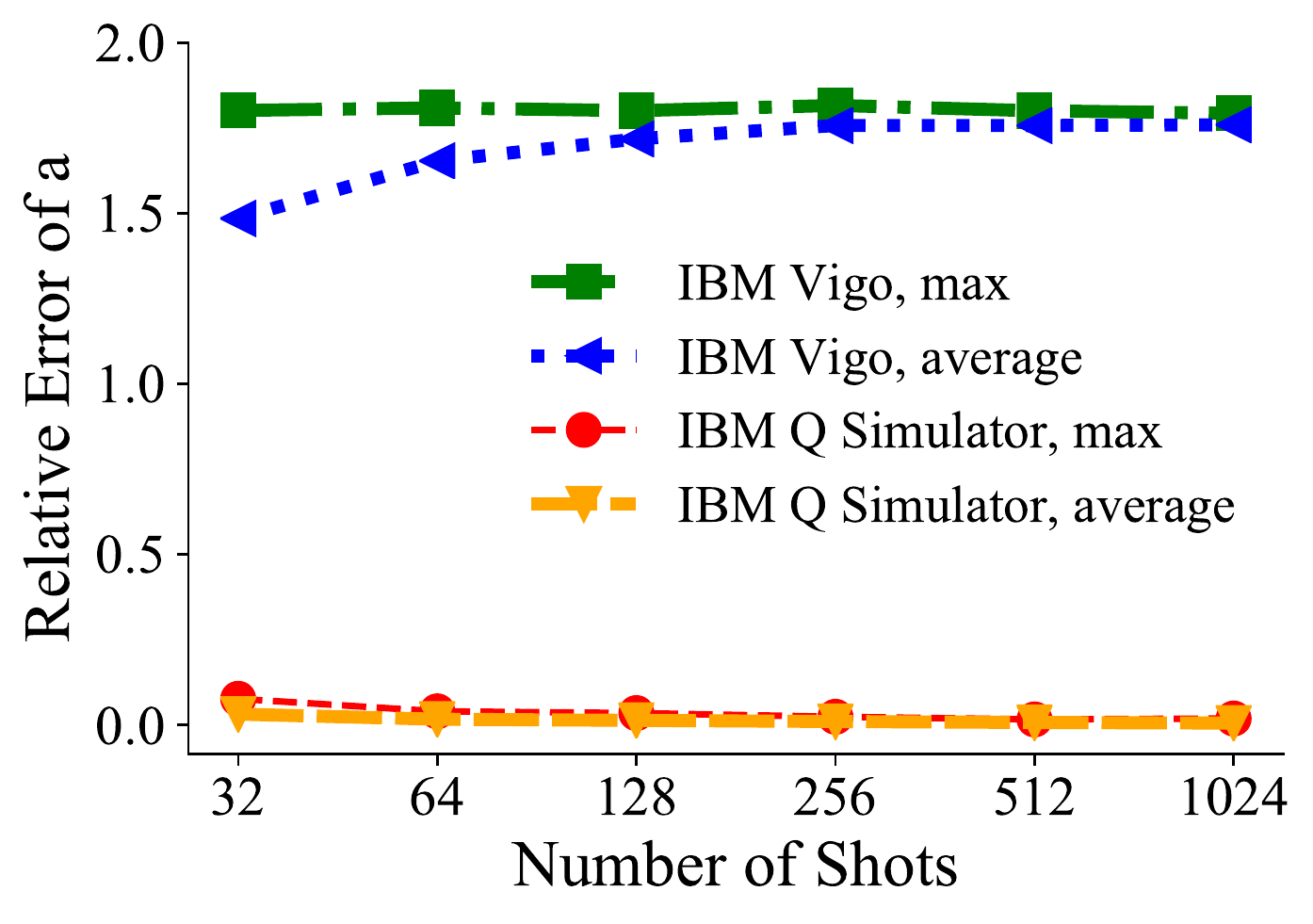}}
  \qquad
  \subfloat[IQAE]{
    \includegraphics[scale=0.5]{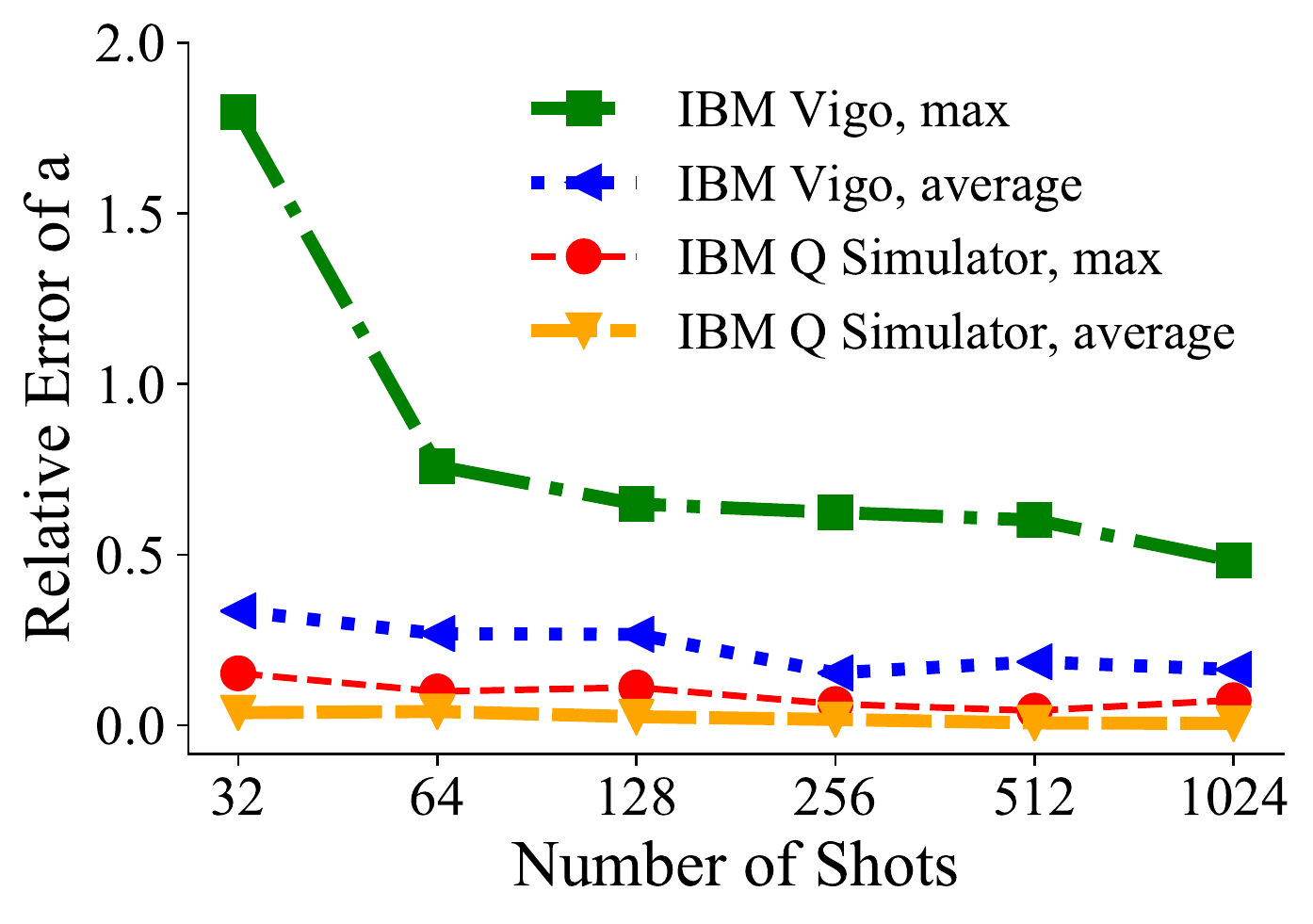}
  }\\
  \vspace{3mm}
  \caption{Relative error for $n=3$-qubit domain for MLQAE and IQAE on IBM Q Vigo and the simulator.}
  \label{fig:MLAE_IAE_rel_a_3q}
\end{figure*}

We present the results from running our simulations with MLQAE and IQAE on IBM Q device, Vigo, and on IBM Q simulator for $2$, $3$, and $10$-qubit domains. The most important parameter for MLQAE is $m$, which corresponds to the number of applications of $\textbf{Q}$. For our MLQAE simulations, we have set this parameter to be $3$, which is the maximum possible value that can be calculated successfully on the device. Increasing $m$ from $3$ to $4$ causes the circuit length to be long enough that the results become noisy on the quantum device because of decoherence and gate errors. For IQAE, the most important parameter is the error tolerance, $\epsilon$, which determines $k$, the power of $\textbf{Q}$. The lower the value of the error tolerance, the more computationally intensive the solution would be. In this study, we have taken this value to be $\epsilon = 0.01$, which was guided by practical considerations, striking a balance between accuracy and feasibility. For each of these two methods, we have used the number of shots ranging from $2^4$ to $2^{10}$, and each run corresponding to a shot number is repeated $30$ times, which we refer to as the number of trials. This procedure leads to a statistically converged value for estimated $a$, the amplitude of the ``good state". 

In the implementation of this experiment, the operator $\mathcal{A}$ and $\mathcal{A}^{-1}$ have the role of a quantum oracle because they compute the function value. Hence, we call $\mathcal{A}$ and $\mathcal{A}^{-1}$ the oracle.
Besides the accuracy itself, the number of oracle calls required to achieve that accuracy is a very important criterion for determining the merits of a method from a practical perspective. The number of oracle calls for MLQAE are completely determined by $m$ and the number of shots, while for IQAE, the number of applications of the oracle is dynamic and depends on our choice of the error tolerance. The more stringent error tolerance leads to more applications of the oracle (or operator $\textbf{Q}$), thus making it computationally expensive.

The main quantities of interest in the study are the estimated $a$, relative error of $a$ and number of oracle calls. To understand the behavior of the two aforementioned quantum amplitude estimation algorithms, we present results from our study in this section.
To estimate the relative error of the estimated $a$, we need to discuss the true value of $a$.
As discussed in Sec.~\ref{sec:implementation}, our experiments compute the following definite integral: 
\begin{equation}
\label{eq:int_pi_over_4_2}
\frac{1}{\pi/4} \int_{0}^{\pi/4}  \sin^2 x ~ dx = \Big(\frac{\pi}{8} - \frac{1}{4} \Big) \Big/ \Big(\pi/4 \Big) = \frac{1}{2} - \frac{1}{\pi} \approx 0.181690 
\end{equation}
by the midpoint Riemann sum.
In the 2-qubit and 3-qubit experiments, the true values of $a$ are $0.179636$ from Eq. (\ref{eq:int_pi_over_4_2q_compute}) and $0.181178$ from Eq. (\ref{eq:int_pi_over_4_3q_compute}), respectively.
For the 10-qubit experiment, we assume $0.181690$ in Eq.~(\ref{eq:int_pi_over_4_2}) as the true $a$ because the Riemann sum of $1024$ discretizations is sufficiently close to the actual value of the integral (the limit of the Riemann sum).

\begin{figure*}[t]
\centering
  \subfloat[MLQAE]{
    \includegraphics[scale=0.5]{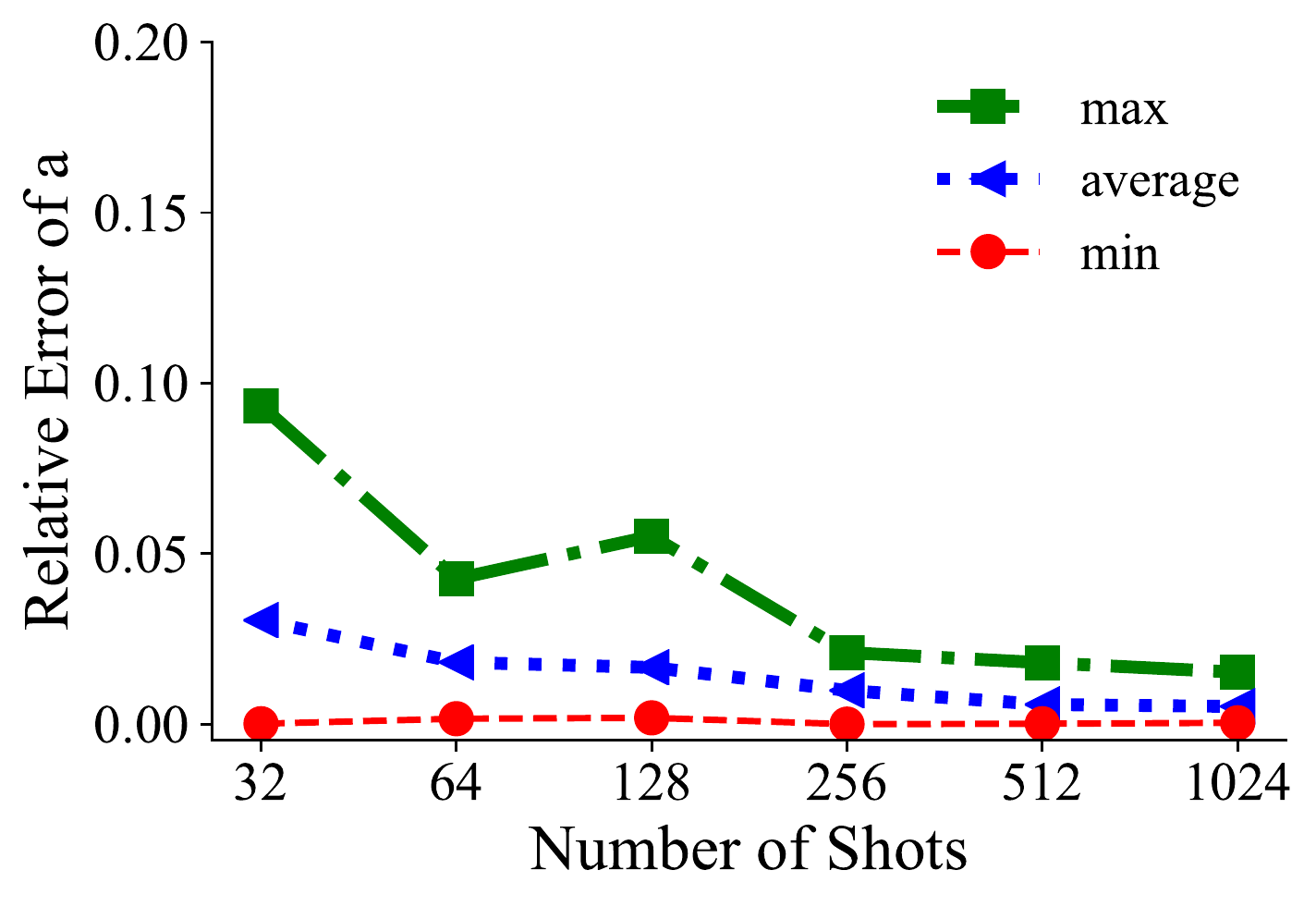}}
  \subfloat[IQAE]{
    \includegraphics[scale=0.5]{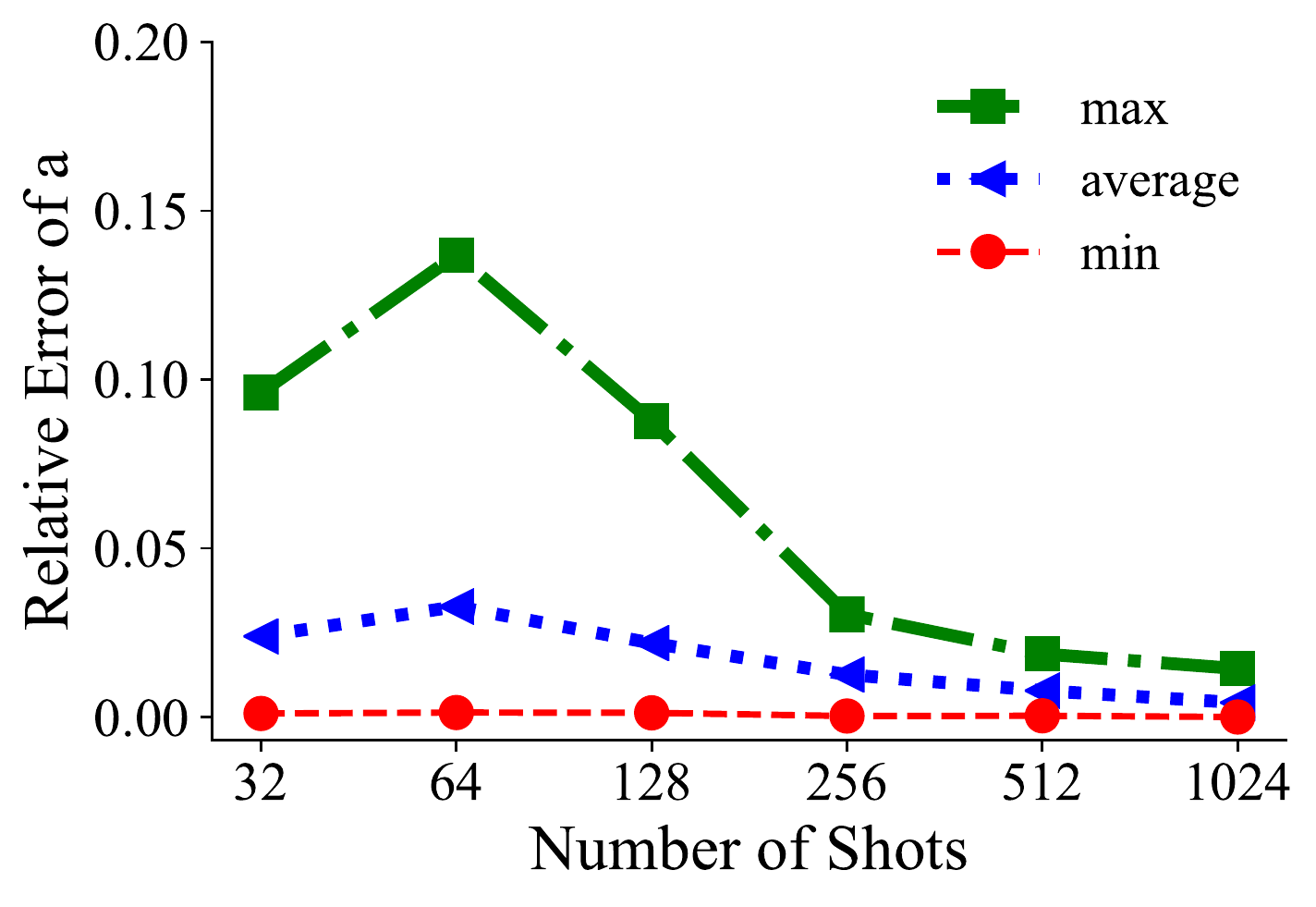}
  }\\
  \caption{Relative error for $n=10$-qubit domain for MLQAE and IQAE on IBM Q simulator.}
  \label{fig:MLAE_IAE_rel_a_10q}
\end{figure*}

\subsection{Maximum Likelihood Quantum Amplitude Estimation}
MLQAE has been run with $m=3$, so the circuits considered here are $\mathcal{A} \ket{0}_{n+1}$, $\textbf{Q} \mathcal{A} \ket{0}_{n+1}$, $\textbf{Q}^2 \mathcal{A} \ket{0}_{n+1}$, $\textbf{Q}^4 \mathcal{A} \ket{0}_{n+1}$. We plot the average and maximum values of the relative error of $a$ for the IBM Q device and the simulator in Fig.~\ref{fig:MLAE_IAE_rel_a_2q}(a). These values are obtained from doing various runs with differing number of shots for a 2-qubit domain. The difference between the quantum device and the simulator runs is significant and is attributed to the quantum device errors. 
A similar trend is observed for the $n=3$ case in Fig.~\ref{fig:MLAE_IAE_rel_a_3q}. The $n=10$ case in Fig.~\ref{fig:MLAE_IAE_rel_a_10q} is run on IBM Q Simulator and gives very accurate results for both MLQAE and IQAE. Comparing the simulator results for the three domain sizes, we see that the relative error is quite comparable for the $n=2$ and $3$ ($\sim 10\%$) cases, but is significantly smaller for the $n=10$ case, especially for a higher number of samples, where it is almost a decade smaller. In Fig.~\ref{fig:MLAE_a_2q} (Appendix), we compare the raw values of estimated $a$ from the simulator and the device. 
The value obtained from the simulator, Fig.~\ref{fig:MLAE_a_2q}(a), coincides pretty well with the true value of $a$, but the average value of estimated $a$ from the device, Fig.~\ref{fig:MLAE_a_2q}(b), differs from the true value by a factor of $2$. 
A similar theme follows for the $n=3$-qubit case for the device runs shown in Fig.~\ref{fig:MLAE_a_3q}, with even larger deviation between the true value and the estimated value on the device. The estimated value for the $10$-qubit case shown in Fig.~\ref{fig:MLAE_IAE_a_10q} coincides almost perfectly with the true value.

\begin{figure*}[t]
\centering
  \subfloat[$n=2$]{
    \includegraphics[scale=0.5]{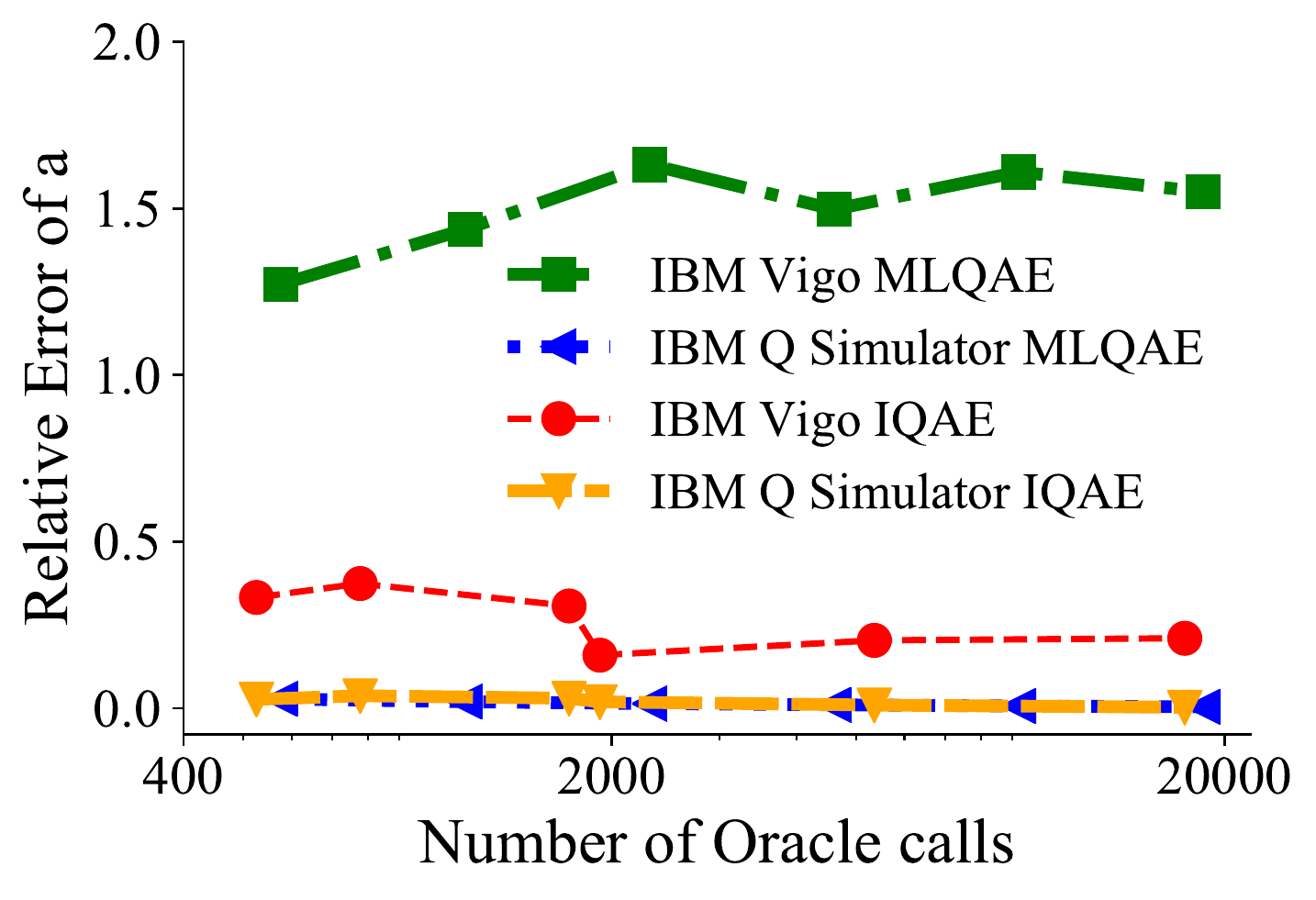}}
  \subfloat[$n=3$]{
    \includegraphics[scale=0.5]{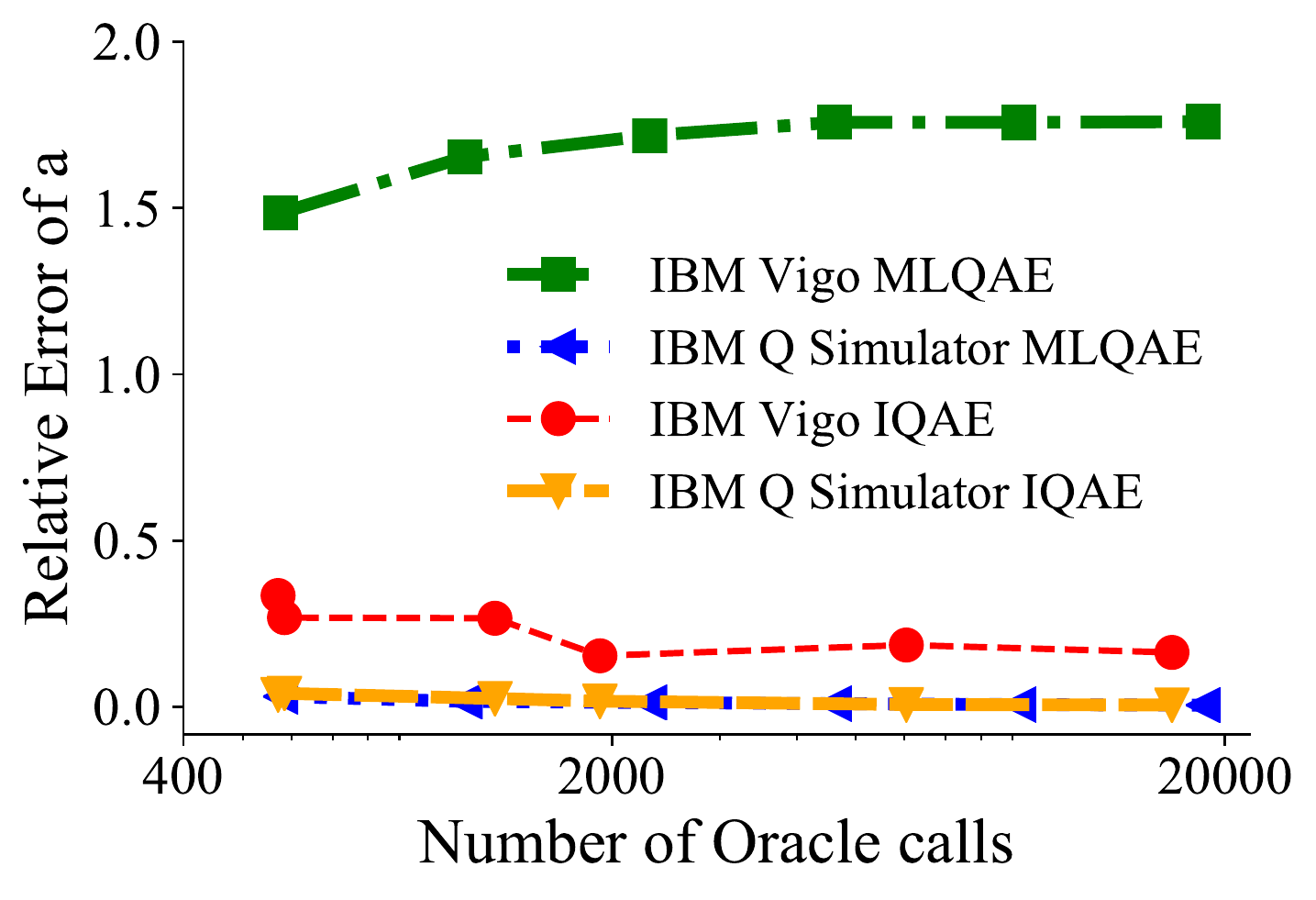}
  }\\
  \caption{The relative error of the estimated $a$ vs. the number of oracle calls, (a) n=2 case, (b) n=3 case}
  \label{fig:MLAE_IAE_oracle_2q3q}
\end{figure*}

\begin{figure*}[t]
\centering
  \subfloat{
    \includegraphics[scale=0.50]{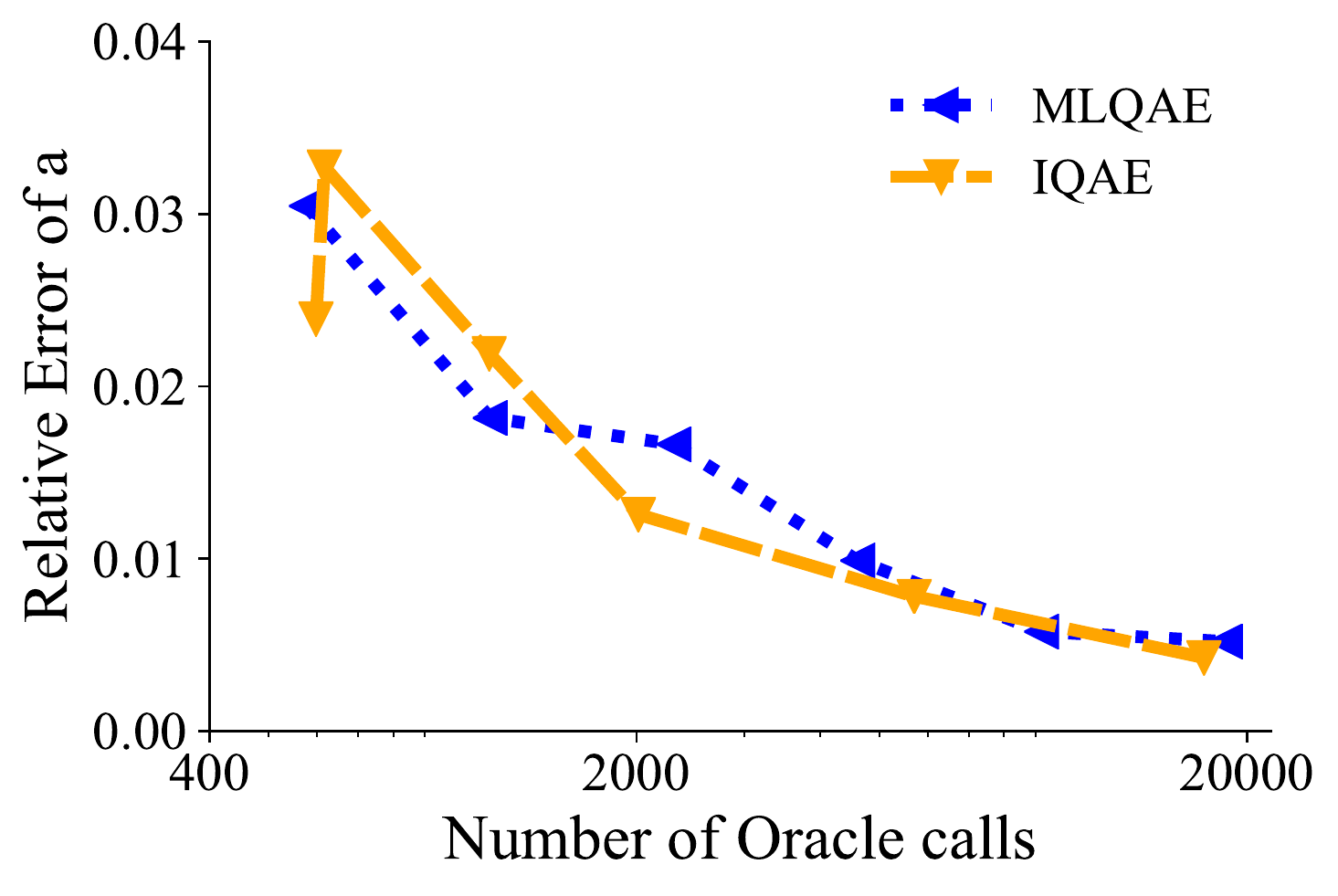}}
  
  \caption{ The relative error of the estimated $a$ vs. the number of oracle calls for $n=10$-qubit domain run on the IBM Q Simulator.}
  \label{fig:MLAE_IAE_oracle_10q}
\end{figure*}

\subsection{Iterative Quantum Amplitude Estimation}
In Fig.~\ref{fig:MLAE_IAE_rel_a_2q}(b), we observe that the relative error of IQAE for the $2$-qubit quantum device run decreases as the number of shots increase. The average simulator results are almost $3$ times less noisy than the device runs on Vigo. This difference is much bigger for the worst case scenario, that is, for the maximum relative error. For $n=3$ in Fig.~\ref{fig:MLAE_IAE_rel_a_3q}, the relative error for IQAE is bigger than that of the $n=2$ case because the increased circuit length for the former leads to more quantum device errors. Similar to IQAE for $n=2$, the relative error for $n=3$ steadily decreases as the number of shots increase. 
In Fig.~\ref{fig:MLAE_IAE_rel_a_10q}(b), the results for running the $n=10$-qubit case on the simulator are presented. The relative error for both methods is very low ($<5\%$) and comparable. In Fig.~\ref{fig:IAE_a_2q} (Appendix), the estimated value of $a$ is compared to the true value for the simulator as well as for IBM Q Vigo. Figure~\ref{fig:IAE_a_3q} also compares the same quantity for $n=3$. The results from the quantum device are much noisier compared to the simulator and the $2$-qubit results. Lastly, in Fig.~\ref{fig:MLAE_IAE_oracle_2q3q}(b), the effect of oracle calls on the relative error can be observed on the device as well as on the simulator for the $2$ and $3$-qubit cases. In the same vein, for the $n=10$-qubit case, the number of oracle calls plotted against the relative error on the simulator is depicted in Fig.~\ref{fig:MLAE_IAE_oracle_10q} and hence, does not include any quantum effects. From this, we observe that both IQAE and MLQAE require a similar number of oracle calls to achieve the same accuracy if quantum effects are ignored. 
\section{Conclusion}
We have compared the implementation and performance of two quantum amplitude estimation algorithms on the basis of the oracle calls, accuracy and efficiency on the NISQ era quantum devices. We have also performed runs on IBM's quantum simulator to highlight the effect of quantum noise on the simulations. From the perspective of implementation on the quantum devices, IQAE seems to perform better as shown in Figs.~\ref{fig:MLAE_IAE_rel_a_2q}, \ref{fig:MLAE_IAE_rel_a_3q}, and \ref{fig:MLAE_IAE_oracle_2q3q}. 
The main reason for the inaccuracy of MLQAE on the quantum device is that the fourth circuit, ($\textbf{Q}^4 \mathcal{A} \ket{0}_{n+1}$), has long circuit depth, which is proportional to the number of $\textbf{Q}$, while the IQAE runs have at most $\textbf{Q}^2$ even though they have more iterations than four.
When we tested IQAE with error tolerance $0.005$ on devices, the runs readily needed $\textbf{Q}^4$ and even $\textbf{Q}^7$. Thus, the results was so noisy that they did not have meaning.
However, in the absence of quantum noise, MLQAE with $m=3$ and IQAE with $0.01$ error tolerance show similar performance as shown in Fig.~\ref{fig:MLAE_IAE_oracle_10q}.

\bibliography{report} 
\bibliographystyle{spiebib} 

\newpage

\appendix

\section{Plots}
\label{app:graphs} 

$\newline$
$\newline$
$\newline$

\begin{figure*}[h!]
\centering
  \subfloat[IBM Q Simulator]{
    \includegraphics[scale=0.5]{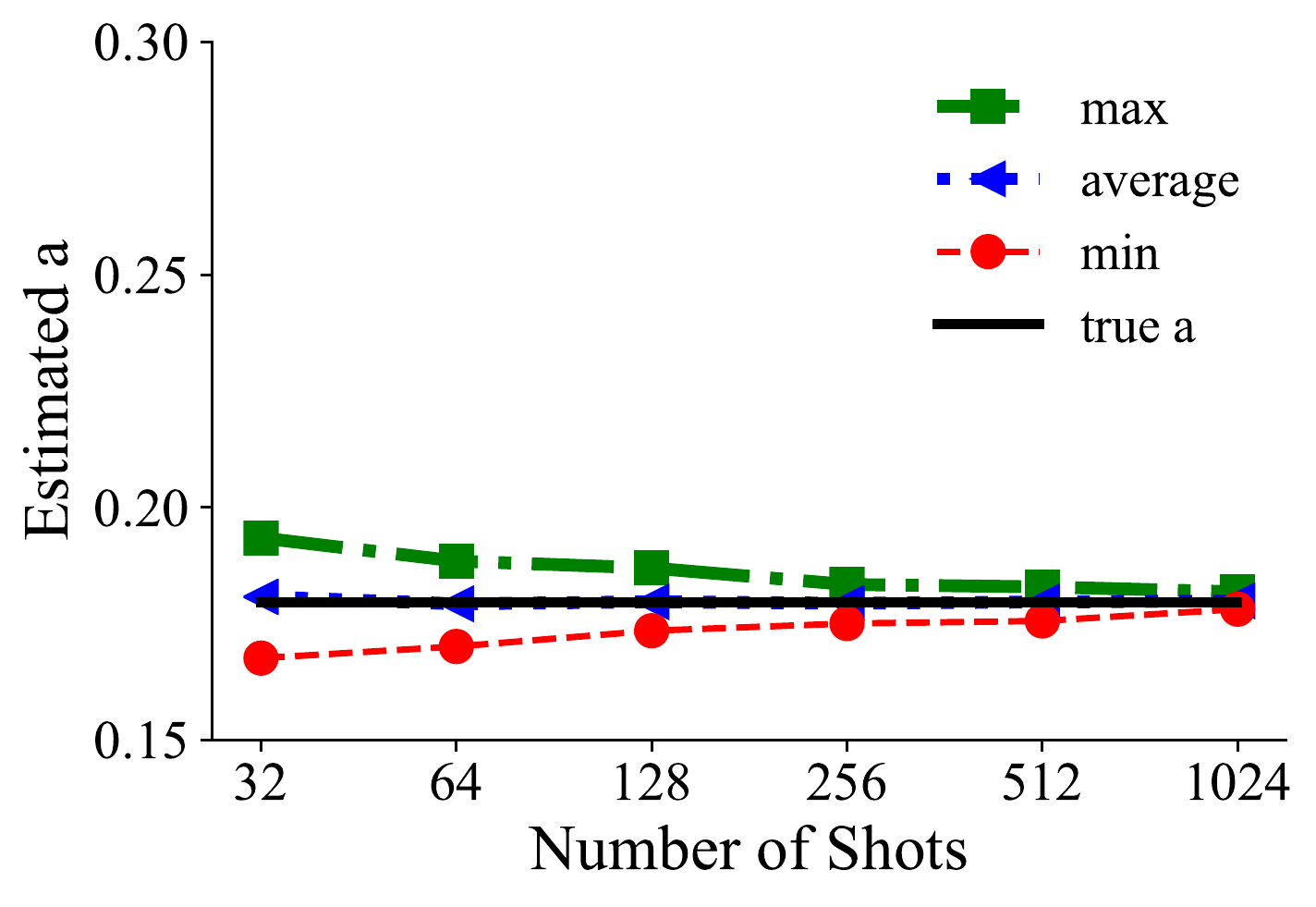}}
  \subfloat[IBM Q Vigo]{
    \includegraphics[scale=0.5]{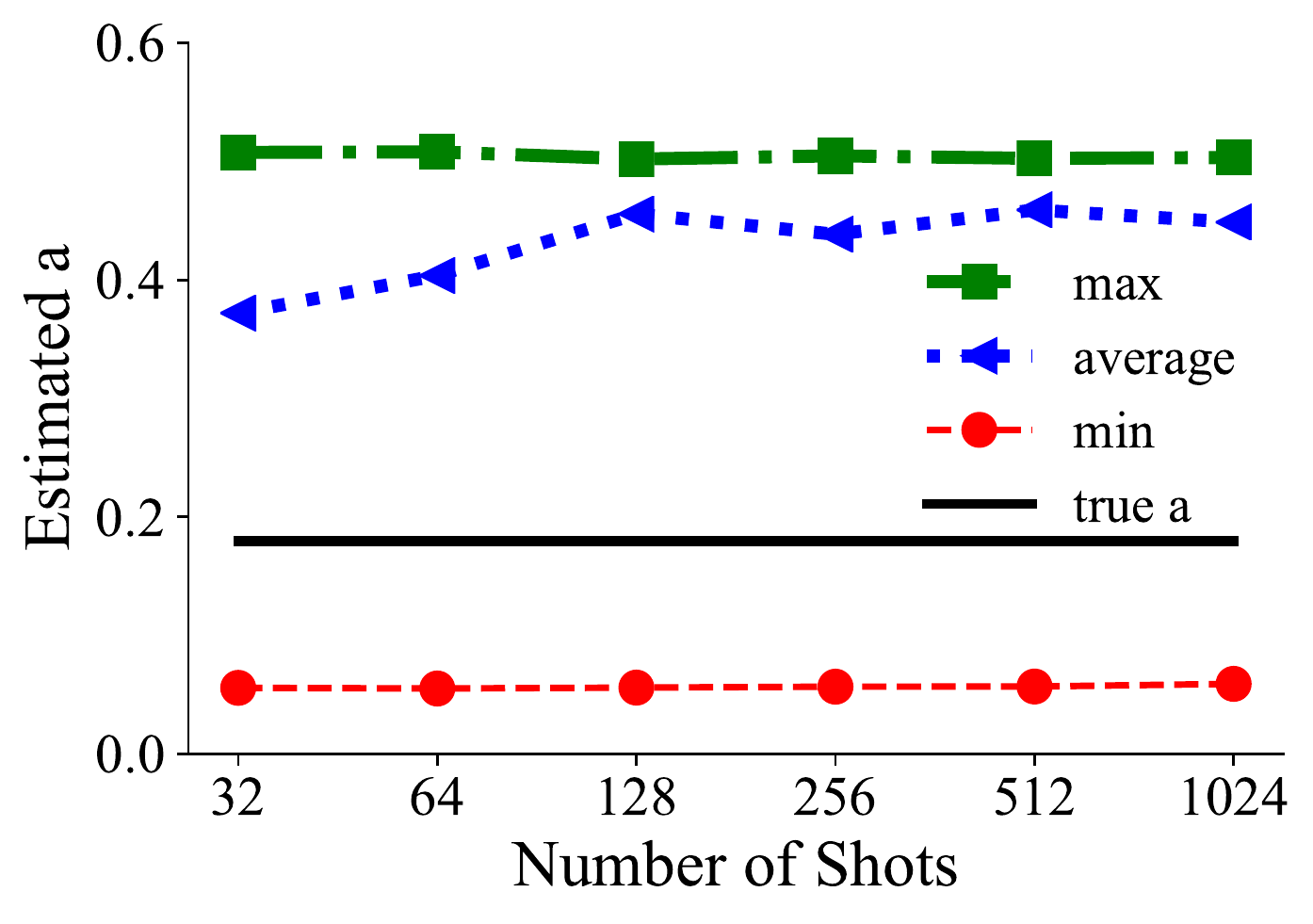}
  }\\
  \vspace{5mm}
  \caption{Estimated $a$ compared to true $a=0.179636$ for $n=2$-qubit domain for MLQAE on IBM Q Simulator and Vigo.}
  \label{fig:MLAE_a_2q}
\end{figure*}

$\newline$
$\newline$
$\newline$

\begin{figure*}[h!]
\centering
  \subfloat[IBM Q Simulator]{
    \includegraphics[scale=0.5]{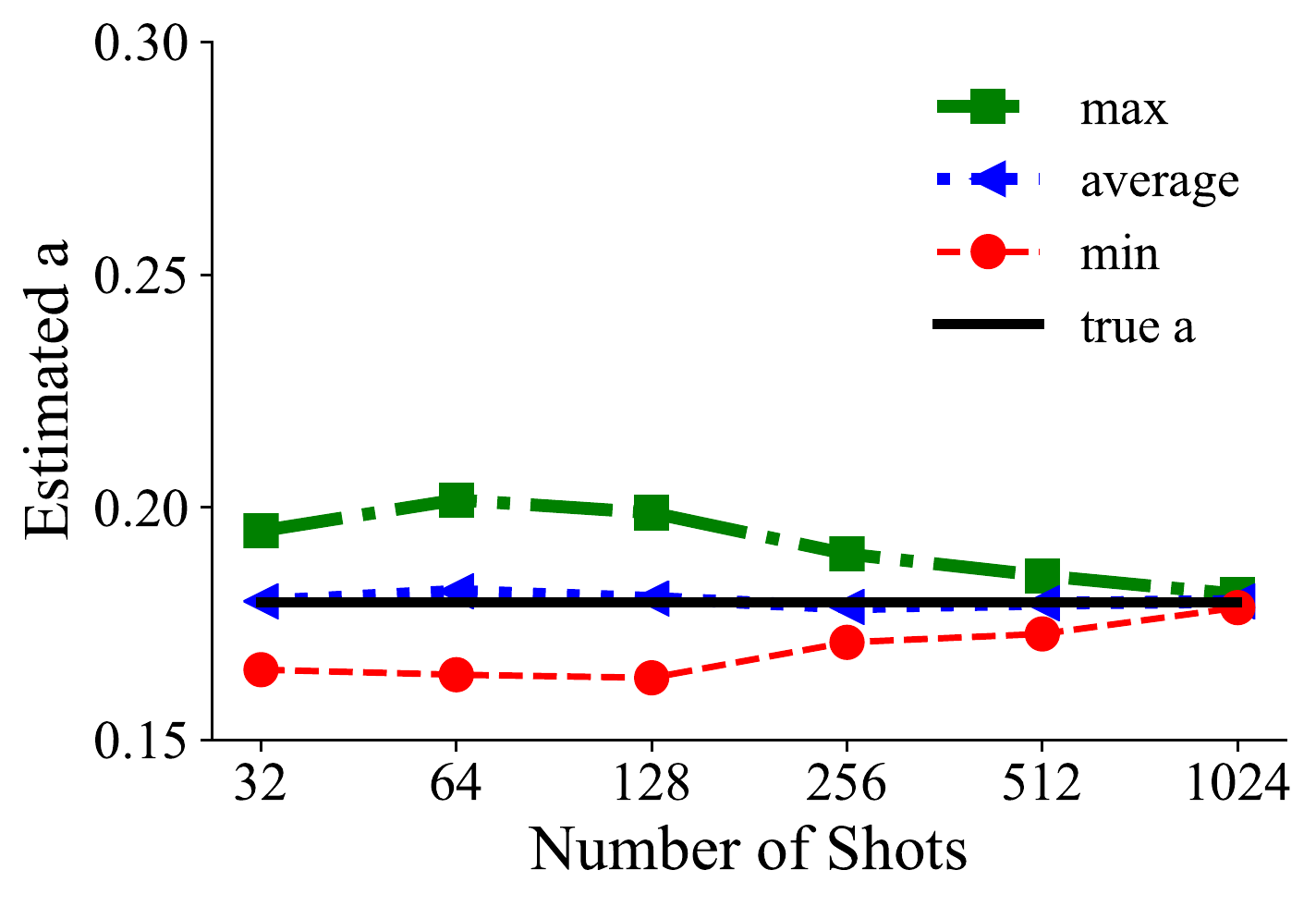}}
  \subfloat[IBM Q Vigo]{
    \includegraphics[scale=0.5]{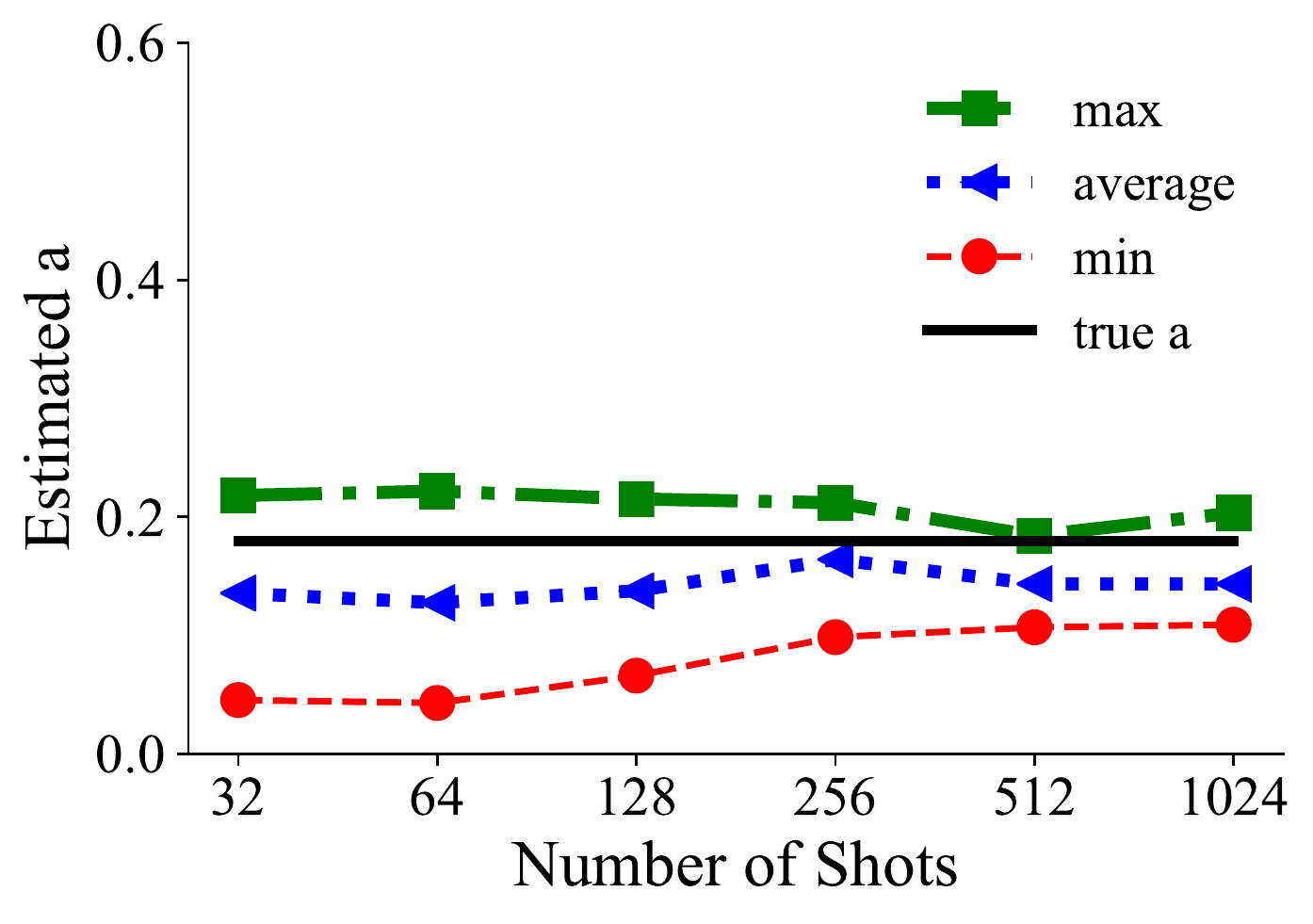}
  }\\
  \vspace{5mm}
  \caption{Estimated $a$ compared to true $a=0.179636$ for $n=2$-qubit domain for IQAE on IBM Q Simulator and Vigo.}
  \label{fig:IAE_a_2q}
\end{figure*}

\begin{figure*}[h!]
\centering
  \subfloat[IBM Q Simulator]{
    \includegraphics[scale=0.5]{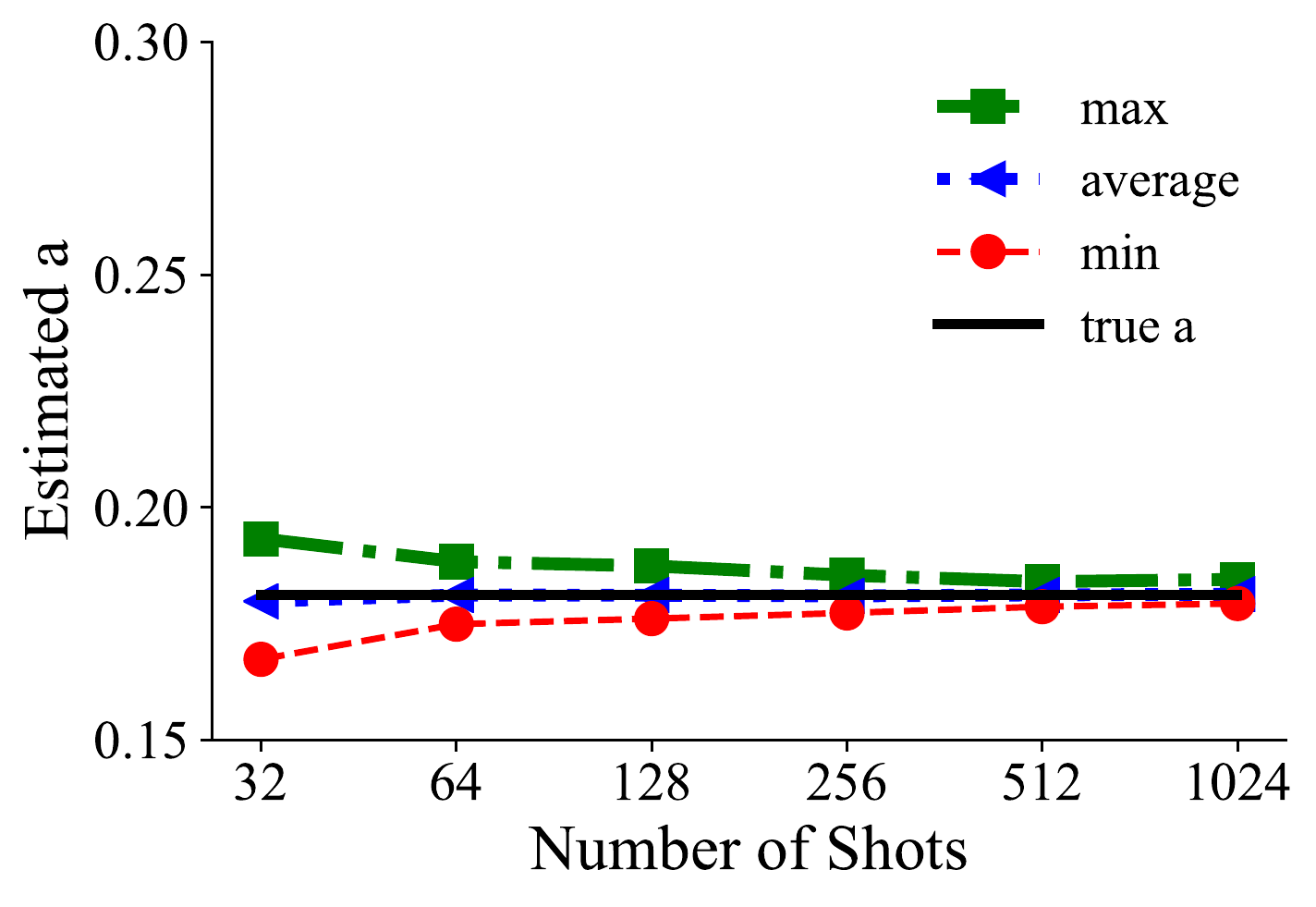}}
  \subfloat[IBM Q Vigo]{
    \includegraphics[scale=0.5]{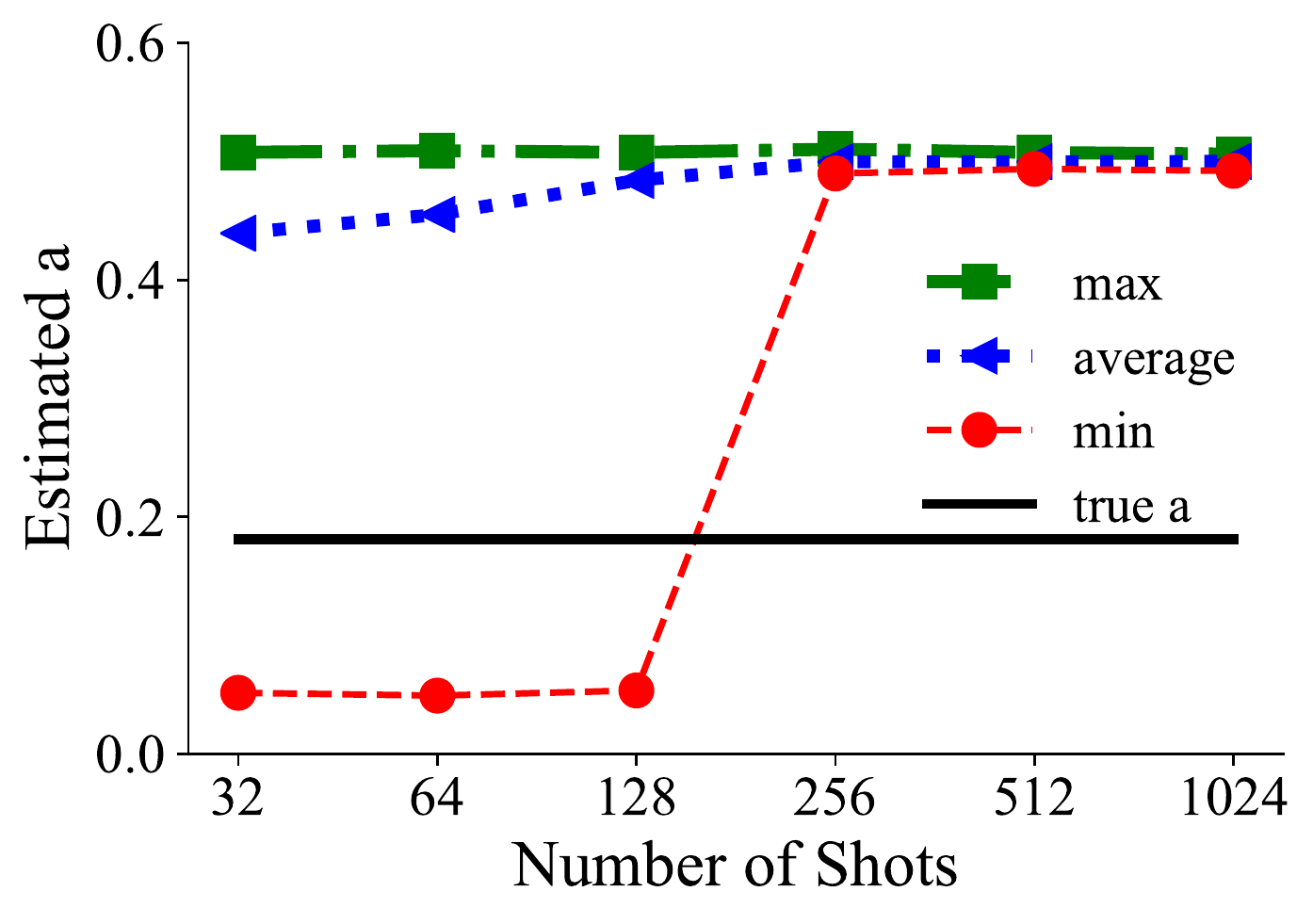}
  }\\
  \vspace{5mm}
  \caption{Estimated $a$ compared to true $a=0.181178$ for $n=3$-qubit domain for MLQAE on IBM Q Simulator and Vigo.}
  \label{fig:MLAE_a_3q}
\end{figure*}

\begin{figure*}[h!]
\centering
  \subfloat[IBM Q Simulator]{
    \includegraphics[scale=0.5]{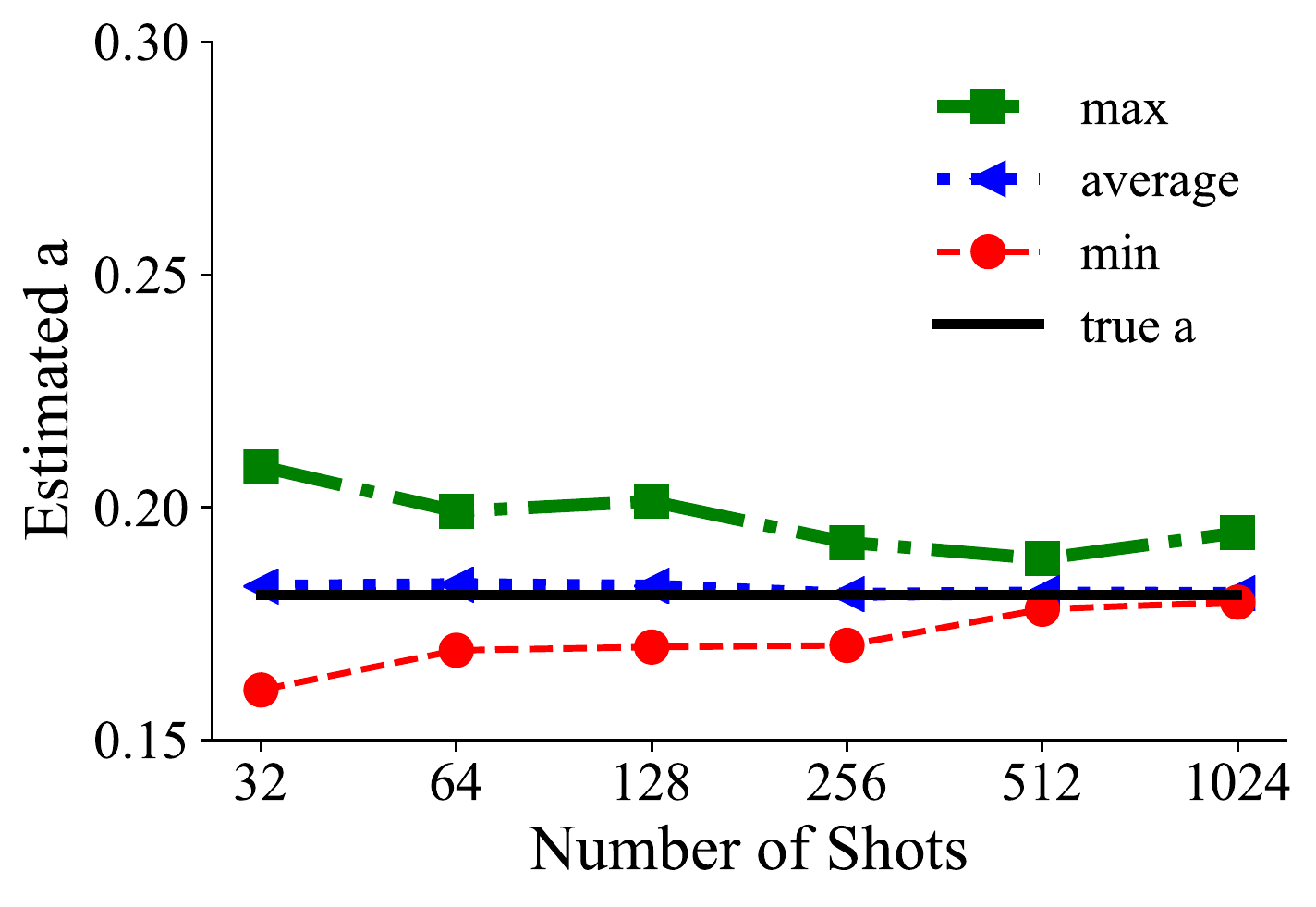}}
  \subfloat[IBM Q Vigo]{
    \includegraphics[scale=0.5]{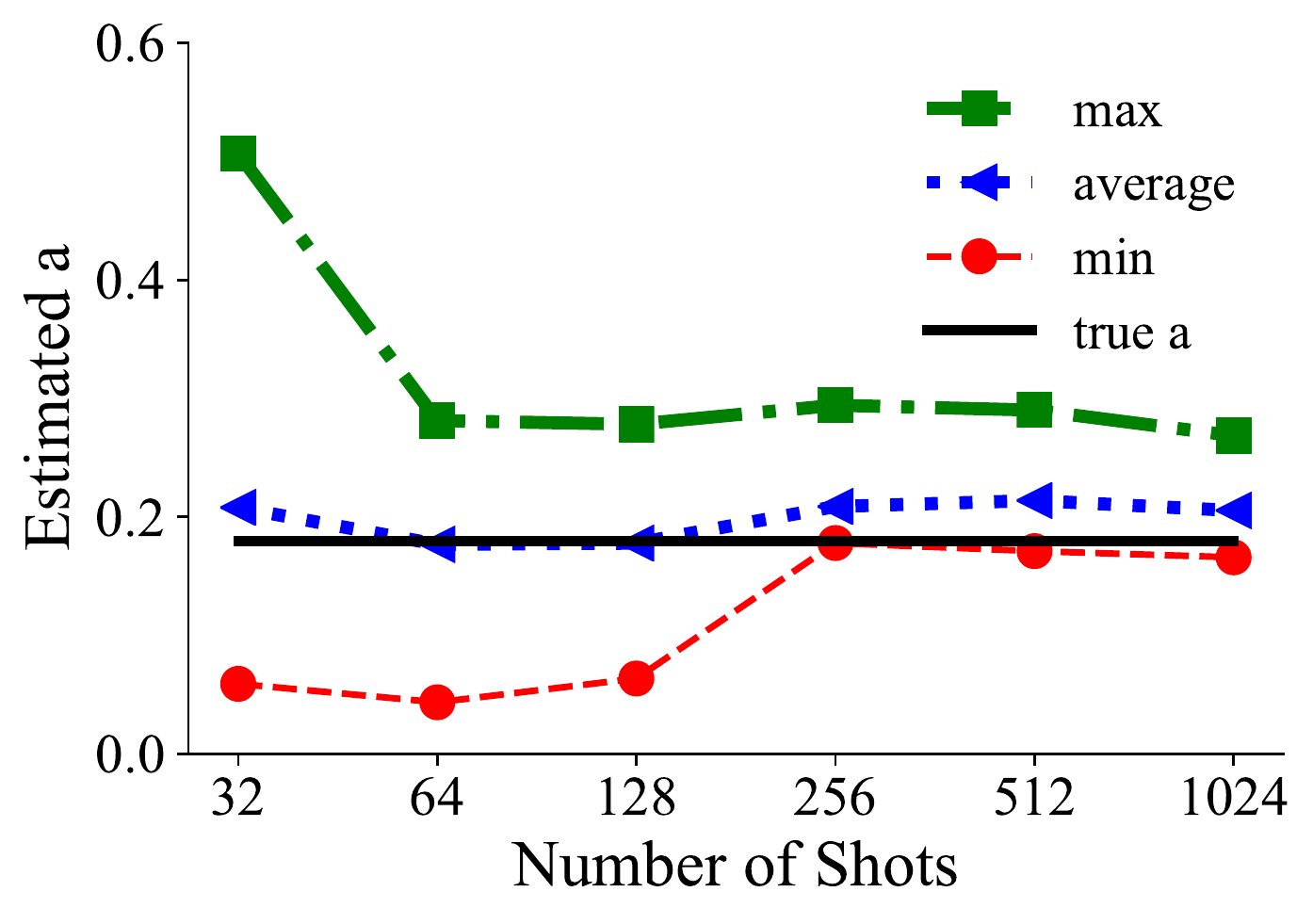}
  }\\
  \vspace{5mm}
  \caption{Estimated $a$ compared to true $a=0.181178$ for $n=3$-qubit domain for IQAE on IBM Q Simulator and Vigo.}
  \label{fig:IAE_a_3q}
\end{figure*}

\begin{figure*}[h!]
\centering
  \subfloat[MLQAE]{
    \includegraphics[scale=0.5]{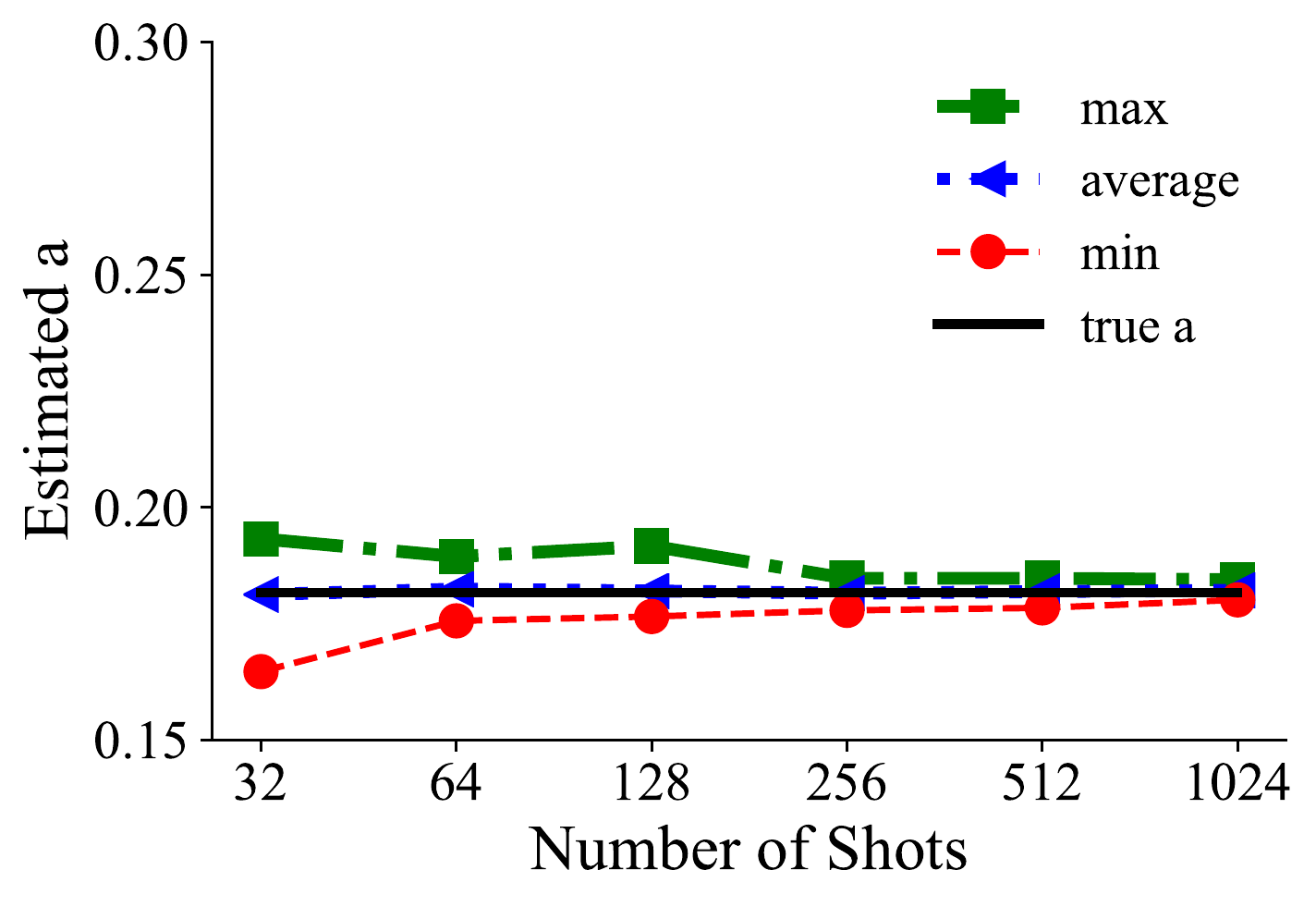}}
  \subfloat[IQAE]{
    \includegraphics[scale=0.5]{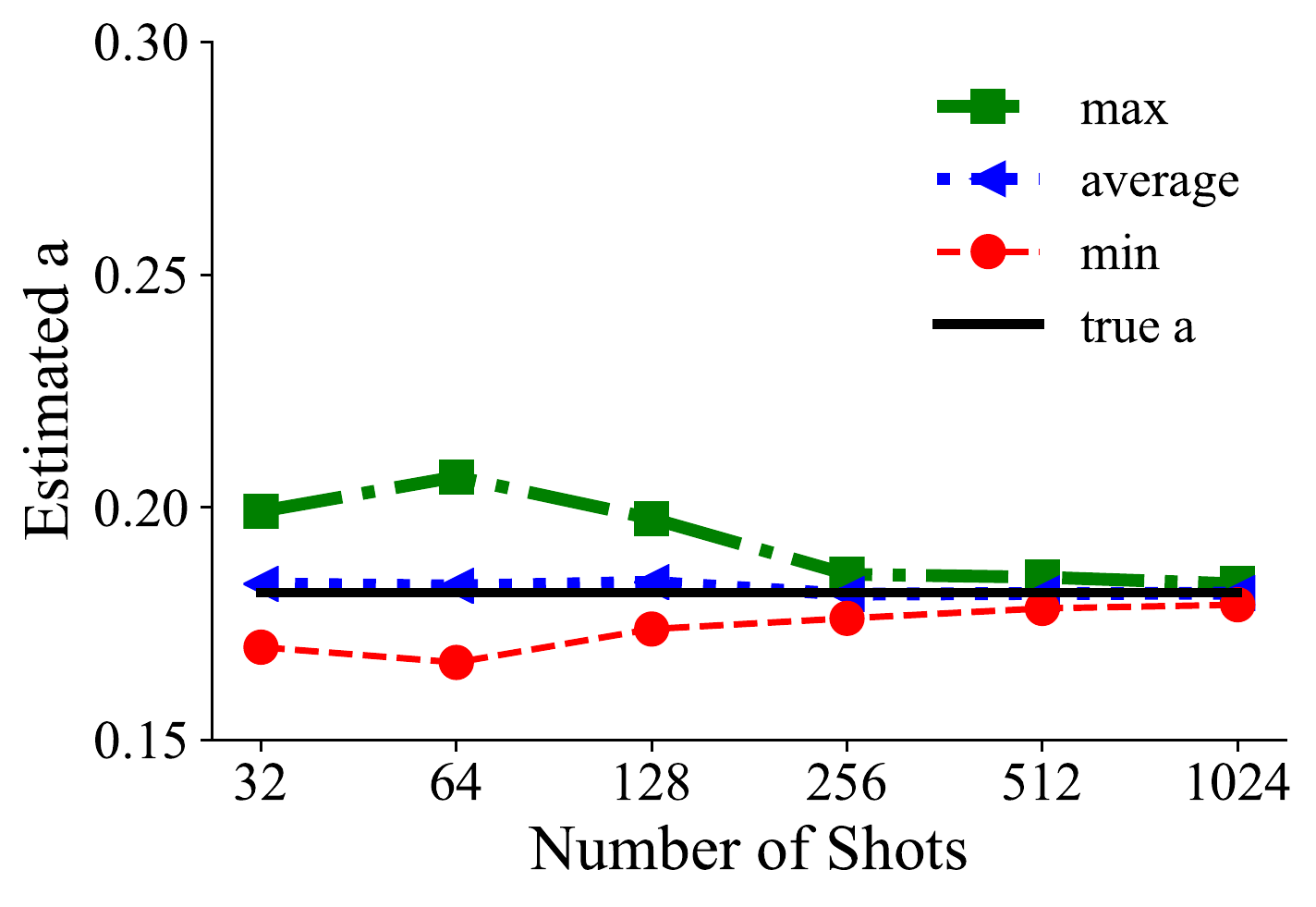}
  }\\
  \vspace{5mm}
    \caption{Estimated $a$ compared to true $a=0.18169$ for $n=10$-qubit domain for MLQAE and IQAE on IBM Q Simulator.}
  \label{fig:MLAE_IAE_a_10q}
\end{figure*}

\end{document}